\documentclass[aps,pra,unsortedaddress, twocolumn,a4paper,amsmath,amssymb,showpacs,%
floatfix,]{revtex4}

\usepackage{graphicx}
\usepackage{epstopdf}

\usepackage[usenames,dvipsnames]{color}

\newcommand{\ket}[1]{\vert{#1}\rangle}

\begin{document}
%
%
%
\title{Controlling vibrational cooling with Zero-Width Resonances: \\An adiabatic Floquet approach}

\date{\today} 

\author{Arnaud Leclerc}\email{Arnaud.Leclerc@univ-lorraine.fr}
\affiliation{Universit\'e de Lorraine, UMR CNRS 7565 SRSMC, 1 Bd. Arago, 57070 Metz, France}
\author{David Viennot}
\affiliation{Institut UTINAM UMR CNRS 6213, Observatoire de Besan\c{c}on,41 bis Avenue de l'Observatoire 25010 Besan\c{c}on Cedex, France}
\author{Georges Jolicard}
\affiliation{Institut UTINAM UMR CNRS 6213, Observatoire de Besan\c{c}on,41 bis Avenue de l'Observatoire 25010 Besan\c{c}on Cedex, France}
\author{Roland Lefebvre}
\affiliation{ISMO, Univ. Paris-Sud, CNRS, Universit\'e Paris-Saclay, 91405 Orsay cedex, France}
\author{Osman Atabek}\email{osman.atabek@u-psud.fr}
\affiliation{ISMO, Univ. Paris-Sud, CNRS, Universit\'e Paris-Saclay, 91405 Orsay cedex, France}

\begin{abstract}
In molecular photodissociation, some specific combinations of laser parameters (wavelength and intensity) lead to unexpected Zero-Width Resonances (ZWR), with in principle infinite lifetimes. Their interest in inducing basic quenching mechanisms have recently been devised in the laser control of vibrational cooling through filtration strategies [O. Atabek et al., Phys. Rev. A87, 031403(R) (2013)]. A full quantum adiabatic control theory based on the adiabatic Floquet Hamiltonian is developed to show how a laser pulse could be envelop-shaped and frequency-chirped so as to protect a given initial vibrational state against dissociation, taking advantage from its continuous transport on the corresponding ZWR, all along the pulse duration. As compared with previous control scenarios actually suffering from non-adiabatic contamination, drastically different and much more efficient filtration goals are achieved. A semiclassical analysis  helps in finding and interpreting a complete map of ZWRs in the laser parameter plane. In addition, the choice of a given ZWR path, among the complete series identified by the semiclassical approach, amounts to be crucial for the cooling scheme, targeting a single vibrational state population left at the end of the pulse, while all others have almost completely decayed. The illustrative example, offering the potentiality to be transposed to other diatomics, is Na$_2$ prepared by photoassociation in vibrationally hot but 
translationally and rotationally cold states.
\end{abstract}

\maketitle




\section{Introduction}
\label{sec:intro}
Quantum control is aimed at designing external pulses in order to achieve efficient transfers between the states of the quantum system under study \cite{rice,tannorbook,shapiro}. This task is crucial in atomic and molecular physics, and has many applications extending from photochemistry to quantum computation. Quantum control has attracted attention among the physics and chemistry communities \cite{kosloff}, but also in applied mathematics for the development of new theoretical methods. 
In strong field molecular physics, antagonistic basic mechanisms like bond softening versus vibrational trapping \cite{Guisti_PRL, Bucksbaum_PRL}, or barrier lowering versus dynamical dissociation quenching \cite{Chateauneuf} have been referred to in the control scenarios of molecular reactivity or even for alignment/orientation purpose \cite{numico}. An even more unexpected quenching mechanism is in relation with the so-called Zero-Width Resonances (ZWR). For certain couples of laser parameters (wavelength and intensity) the photodissociation rate vanishes leading to, in principle, infinitely long lived resonances (ZWRs) in both continuous wave \cite{chrysos, atabeklefebvre} and pulsed regimes \cite{catherine}. In a diatomic molecule, a destructive interference is in play taking place among fluxes contributing to the outgoing scattering amplitude of an initial vibrational state decaying through two laser-induced adiabatic channels. When dealing with intramolecular couplings, the specific requirement for such destructive interference can only be fulfilled for some particular interchannel coupling amplitudes. Despite strong interchannel coupling, the observation of some narrow rotational lines in IBr predissociation provides such an example \cite{Child2}. Radiative interactions offer the possibility of a continuous tuning of laser parameters to reach a ZWR.

Our laser control objective is the vibrational cooling of diatomic molecules produced by photoassociation in translationally and rotationally cold, tightly bound states \cite{fatemi}. 
The filtration strategy we are referring to aims at protecting against photodissociation a given single vibrational state, while all others are decaying fast. This is achieved through an adiabatic and continuous transport of this state population on its corresponding ZWR all along a laser pulse adequately shaped and frequency chirped. Starting from a given vibrational distribution, the efficiency of such a control remains on 
two crucial issues: (i) The best possible protection of the given vibrational state; (ii) the highest decay rates for all the others. The first issue deals with the adiabatic transport model. Recently, we built an intuitive transport scheme by chirping a laser pulse such as to merely combine, at each time within the pulse duration, the wavelength and the intensity corresponding to the following of a ZWR originating, in field-free conditions, from the vibrational state we wish to protect. Although this simple scheme does not result from a formal adiabatic treatment, 
we obtained encouraging results in the case of a light diatomic like H$_2^+$, with 70$\%$ of the initial population left at the end of the pulse 
thanks to the relatively slowly varying pulse envelope and modest chirp amplitude \cite{AtabekRC}. 
It turns out that the situation is drastically different for Na$_2$ where almost the whole initial population can photodissociate. This prompted us to go through a full adiabatic transport control scheme based on the adiabatic Floquet theory \cite{guerin} completely reformulated to account for ZWRs in a pulsed regime. The second issue concerns the anharmonicity and relatively high density of vibrational levels in heavier species like Na$_2$. The consequence is that laser pulses shaped so as to protect a given vibrational level can have characteristics close to the ones appropriate for at least partly protecting neighboring levels. A semiclassical analysis helps in finding and interpreting, in the laser parameter plane, a complete map of ZWRs originating from a field-free vibrational level. We show how a choice can be done of a particular ZWR path, among the complete series identified by the semiclassical analysis. This aims at
realizing the most efficient filtration, namely the best compromise for selectively protecting the single vibrational state in consideration.

The paper is organized in the following way: 
In section \ref{sec2}, 
ZWRs are introduced 
within a two-state photodissociation model referring both to a time-independent close-coupled Floquet Hamiltonian formalism and a semiclassical interpretation. Two computational methods, either grid or global are presented. Section \ref{sec3} 
is devoted to a complete derivation of the adiabatic control theory. The filtration strategy as applied to Na$_2$ is discussed in Section \ref{sec4}, 
by introducing the model, the ZWR maps, the adiabatic transport dynamics and the choice of the optimal ZWR path.

\section{Zero-Width Resonances}\label{sec2}

In this Section, we examine the rather unexpected property of a resonance state, originating from a bound state coupled to a dissociative continuum, 
to possibly acquire an infinite (or quasi-infinite) lifetime. 
Such exotic resonances have already been discussed in the literature, first in the context of accidentally narrow rotational lines in predissociation \cite{child}, then as Bound Sates in Continuum (BIC) \cite{BIC}, and more recently as Zero-Width Resonances (ZWR) \cite{atabeklefebvre}. In the following, the physical context is molecular photodissociation and our illustrative example is Na$_2$.

\subsection{Photodissociation Model} \label{photodissmodel}
We briefly recall  the model used to describe the photodissociation of a rotationless field-aligned diatomic molecule in a single spatial dimension (the internuclear distance $R$) with only two electronic states labeled $\arrowvert 1 \rangle$ and $\arrowvert 2 \rangle$ in a Born-Oppenheimer approximation. Reduced dimensionality with the frozen rotation assumption is validated by considering linearly polarized light and 
short pulse durations we refer to hereafter, as compared to the rotational periods of the molecular species.
The time-dependent wave function is written as:
\begin{equation}
 \arrowvert \Phi (R,t) \rangle = \vert \phi_1(R,t) \rangle \arrowvert 1 \rangle +
\vert  \phi_2 (R,t) \rangle \arrowvert 2 \rangle.
\end{equation}
The nuclear dynamics is governed by the Time Dependent Schr\"odinger Equation (TDSE) :
\begin{eqnarray}
& i\hbar \frac{\partial}{\partial t} \left[\begin{array}{c} \phi_1 (R,t)  \\
\phi_2 (R,t) \end{array}\right] 
 =   \left( T_N + \left[\begin{array}{c c} V_1(R)&0 \\
0&V_2(R) \end{array}\right] \right. & \nonumber \\ 
& - \left. \mu_{12}(R)\mathcal{E}(t)\left[\begin{array}{c c} 0&1 \\
1&0 \end{array}\right] \right)  \left[\begin{array}{c} \phi_1 (R,t) \\
\phi_2 (R,t) \end{array}\right] & 
\label{TDSE}
\end{eqnarray}
 $T_N$ represents the nuclear kinetic energy operator. $V_1(R)$ and $V_2(R)$ are the Born-Oppenheimer potentials. $\mu_{12}(R)$ is the electronic transition dipole moment between states $\arrowvert 1 \rangle$ and $\arrowvert 2 \rangle$.
 $\mathcal{E}(t)$ is the linearly polarized electric field amplitude.
 We first consider the case of a continuous wave (cw) laser, 
 \begin{equation}
 \mathcal{E}(t) = E\cos(\omega t)
 \end{equation}
 with an intensity ($I\propto E^2$), a frequency $\omega$ and a wavelength $\lambda= 2 \pi c /\omega$, $c$ being the speed of light.
 In this strictly periodic case, the Floquet ansatz being applied:
\begin{eqnarray}
\left[ \begin{array}{c} \phi_1 (R,t) \\
\phi_2 (R,t) \end{array}\right] = e^{-iE_v t/\hbar} \left[ \begin{array}{c} \chi_1 (R,t)  \\
\chi_2 (R,t) \end{array}\right].
\label{eq:anzats}
\end{eqnarray}
Due to the periodicity in time of $\chi_k(R,t)   (k=1,2)$ these functions can be Fourier expanded:
\begin{equation}
\chi_k(R,t)=\sum_{n=-\infty}^{\infty} e^{in\omega t} \varphi_n^k(R)
\label{eq:expansion}
\end{equation}
where the Fourier components satisfy a set of coupled differential equations, for any $n$:
\begin{eqnarray}
&\left[ 
T_N +V_{1,2}(R)+n\hbar \omega-E_v \right]\varphi_{1,2}^n(R)   \nonumber \\
&-1/2 E \mu_{12}(R) \left[\varphi_{2,1}^{n-1}(R) + \varphi_{2,1}^{n+1}(R)\right]=0
\label{eq:closecoupledmulti}
\end{eqnarray}
For moderate field intensities inducing single-photon processes, these equations obviously simplify into two close-coupled equations:
\begin{equation}
\left[ T_N +
V_1(R)+\hbar \omega-E_v \right]\varphi_{1,v}(R)-1/2 E \mu_{12}(R)
\varphi_{2,v}(R)=0 
\nonumber
\end{equation}
\begin{equation}
\left[ T_N +
V_2(R)-E_v \right]\varphi_{2,v}(R)-1/2 E \mu_{12}(R)
\varphi_{1,v}(R)=0   
\label{eq:closecoupled}
\end{equation}
where we have kept only the  Fourier component $(n=1)$ of $\chi_{1,v}(R,t)$ and the zero-frequency Fourier component $(n=0)$ of $\chi_{2,v}(R,t)$, denoted $\varphi_{1,v}(R)$ and $\varphi_{2,v}(R)$, respectively.
It has to be noted that a specific solution ($v$) has been identified in Eq.(\ref{eq:closecoupled}), by labeling both the eigenenergy $E_v$ and the corresponding eigenvector $\chi_{k,v}(R,t)$ and their Fourier components $\varphi_{k,v}(R)$.
Resonances are solutions with Siegert type outgoing-wave boundary conditions \cite{siegert} and have complex quasi-energies of the form 
$\Re(E_v)-i\Gamma_v/2$, where $\Gamma_v$ is the resonance width related to the decay rate.
In the following, the label $v$ designates both the field-free vibrational level and the laser-induced resonance originating from this vibrational state.

Now relaxing the cw laser assumption, we consider 
a chirped laser pulse with parameters $\epsilon(t)\equiv\{E(t), \omega(t)\}$ involving slowly varying envelope and frequency. The molecule, initially in a particular field-free vibrational state $v$, is supposed to be adiabatically driven by such a pulse. Adiabaticity means here that a unique resonance $\Phi_v(t)$, labeled $v$ according to its field-free parent bound state $v$, is followed during the whole dynamics. This resonance wave function involves, through its complete basis set expansion, a combination of both bound and continuum eigenstates of the field-free molecular Hamiltonian. But the important issue is that, at the end of the pulse, the molecule is again on its initial single vibrational state $v$ (adiabaticity condition). For such open systems, contrary to dynamics involving  bound states only, there is unavoidably an irreversible  decay process, precisely due to the fact that vibrational continuum states are temporarily populated under the effect of the pulse. A quantitative measure of such a decay could be given in terms of the overall fraction of non-dissociated molecules, assuming a perfect adiabatic following of the selected resonance
 \cite{atabeklefebvre}:
\begin{equation} \label{P-undiss}
 P_{v}(t) \; = \; \exp\left[- \hbar^{-1}\int_0^{t} \Gamma_v(\epsilon (t'))~dt'\right] \;\; 
\end{equation}
where the decay rate $\Gamma_v(\epsilon(t))$ is associated with the relevant Floquet resonance quasi-energy $E_v(\epsilon(t))$ using the instantaneous field parameters $ \epsilon(t) \equiv\{E(t), \omega(t)\}$ at time $t$.

There are two central control issues: (1) Investigating how rates are changing with the field parameters and in particular find optimal combinations $\epsilon^{\text{ZWR}}(t)$ for which these rates are small enough (or even ideally zero) to insure the survival of the vibrational state $v$ to the laser excitation, that is $P_{v}(\tau) \approx 1$, $\tau$ being the total pulse duration;
(2) Predicting how well the adiabatic following is effectively realized (this point will be analysed in more details in section \ref{sec3}).


\subsection{Semiclassical interpretation}
The occurrence of ZWRs, at least  under some particular circumstances, is rationalized by the semiclassical theory of predissociation of a diatomic molecule \cite{child}. The formalism deals with the so-called adiabatic potentials $V_{\pm}(R)$ resulting from the diagonalization of the molecule-field interaction matrix and predicts dissociation quenching resulting from a null value of the outgoing scattering amplitude in the lower (open) adiabatic channel $V_-$, if the following two conditions are simultaneously fulfilled\cite{Bandrauk,Child2}:
\begin{equation}
	\int_{R_{+}}^{R_0} dR'~k_{+}(R')+\int_{R_{0}}^{R_{t}} dR' k_{+}(R')+ \chi=(\tilde{v}_{+}+\frac{1}{2})\pi
	\label{eq:semiclassical1}
\end{equation}
and
\begin{equation}
	\int_{R_{-}}^{R_0} dR'~k_{-}(R')+\int_{R_{0}}^{R_{t}}dR' k_{+}(R') =(\tilde{v}+\frac{1}{2})\pi
	\label{eq:semiclassical2}
\end{equation}
In  Eqs.(\ref{eq:semiclassical1},\ref{eq:semiclassical2}) the wavenumbers are $k_{\pm}(R)=\hbar^{-1}[2m(\varepsilon-V_{\pm}(R)]^{1/2}$, $m$ is the reduced nuclear mass, $R_{\pm}$ are the left turning points of $V_{\pm}$ potentials, $R_t$ is the right turning point of $V_+$ and $R_0$ is the diabatic crossing point resulting from field-dressing. With integer $\tilde{v}_+$ and $\tilde{v}$ these conditions lead to Bohr-Sommerfeld quantization involving a coincidence between two energies; namely one $\varepsilon=\varepsilon_{\tilde{v}^+}$ of the upper adiabatic potential $V_+(R)$, with a phase correction $\chi$, which in weak coupling is $-\pi/4$ \cite{Child2}, and another $\varepsilon=\varepsilon_{\tilde{v}}$ of a potential made of two branches, namely, $V_-(R)$ for $R \leq R_0$, and $V_+(R)$ otherwise. For a weak coupling, this is practically the diabatic attractive potential $V_1(R)$. More precisely, the coincidence condition is implemented in the expression of the resonance width $\Gamma_v$  \cite{Child2}:
\begin{equation}
	\Gamma_v= \frac{2\pi}{\hbar}  \frac{e^{2\pi\nu} (e^{2\pi\nu}-1) \omega_d \omega_+}{[\omega_++(e^{2\pi\nu}-1)\omega_d]^3} (\varepsilon_{\tilde{v}}-\varepsilon_{\tilde{v}_{+}})^2
	\label{eq:Gammasem}
\end{equation} 
$\omega_d$ and $\omega_+$ are the local energy spacings of the modified diabatic and adiabatic potentials respectively. $\nu$ is the Landau-Zener coupling parameter:
\begin{equation}
	\nu=\frac{\mu_{12}^2(R_0)E^2}{\hbar\bar{v}|\Delta F|}
	\label{eq:Landau-Zener}
\end{equation}
where $\bar{v}$ and $\Delta F$ are the classical velocity and slope difference of the diabatic potentials at $R_0$.
Clearly, the two energies $\varepsilon=\varepsilon_{\tilde{v}}$ and $\varepsilon=\varepsilon_{\tilde{v}^+}$, and therefore the width $\Gamma_v$ are dependent on field parameters, i.e., both frequency (or wavelength) and amplitude (or intensity). This is in particular due to the ($\lambda$, $I$)-dependence of the corresponding field-dressed adiabatic potentials $V_{\pm}(R)$. All other factors building up $\Gamma_v$ in Eq.(\ref{eq:Gammasem}), that is the coupling $\nu$ and the local energy spacings $\omega_d$, $\omega_+$, also depend on field parameters. Contrary to predissociation where such coincidences can only be accidental since there is no easy and continuous way to modify potentials and interstate couplings (electronic or spin-orbit), for a diatomic molecule submitted to an electromagnetic field, a fine tuning of the wavelength and intensity produces them at will. This explains the occurrence of  ZWRs in photodissociation. Moreover, for a wavelength $\lambda$ which roughly brings into coincidence the levels $\tilde{v}$ (corresponding to the field-free vibrational level $v$ in consideration) and $\tilde{v}_+=0$, a fine tuning of the intensity $I$ will result in an accurate determination of a ZWR, that is $\Gamma_v(\lambda , I)=0$. 
In some cases, a stronger field (higher $I$) may also bring into coincidence $\tilde{v}$ with $\tilde{v}_+=1$, producing thus a second ZWR, for the same wavelength, and so on for $\tilde{v}_+=2,3...$ But, one can also envisage slightly different wavelengths which build energetically close enough $\tilde{v}$ and $\tilde{v}_+=0$ levels in a field-dressed picture, such that a subsequent fine tuning of the intensity brings them into precise coincidence. This flexibility offered by the field parameters that, in principle, can be continuously modified, is at the origin of not only quasi-zero width photodissociation resonances, but also for their multiple occurrence in the ($\lambda$, $I$)-parameter plane \cite{multiZWR}.

\subsection{Computational Methods} \label{computmethods}
We are referring to two classes of computational methods for an accurate determination of laser parameters $\epsilon^{\text{ZWR}}(t)$ producing a ZWR: (i) Grid methods and (ii) Global methods. To be more specific, for these calculations, $t$ has not to be considered as a time variable, but rather as a parameter. To a fixed $t$ corresponds a set of field parameters $\epsilon(t)\equiv\{E(t), \omega(t)\}$ or equivalently $\epsilon(t)\equiv\{I(t), \lambda(t)\}$, which determine a specific cw laser, fulfilling the single period requirement of the Floquet ansatz of Eq.(\ref{eq:anzats}). The time-independent system of close-coupled equations for the $R$-dependent Fourier coefficients $\varphi_{1,2}(R)$ of Eqs. (\ref{eq:closecoupled}) are solved for these given field amplitude and frequency.

\subsubsection{Grid methods} \label{gridmethods}

These methods are based on so-called shooting-matching techniques. 
The Fox-Goodwin propagation algorithm \cite{fox} is used in conjunction with a properly chosen set of imposed boundary conditions, which accounts for the correct behavior of the channel wave functions both at short and large distances. The algorithm consists in an iterating sequence of two steps \cite{atabek03}: (i) Shooting: With an initial guess for the quasi-energy $E_v$, properly initialized Fox-Goodwin matrices are constructed at each point of a grid (along $R$) in terms of independent solution matrices propagated inward and outward; (ii) Matching: The criterion for convergence is a condition to be fulfilled for matching both the functions and their derivatives on two adjacent points of the grid, by properly changing the energy $E_v$. While regularity ($\varphi_{1,2}(R\rightarrow 0)=0$) is imposed at the origin for both open and closed channels,
 which are classically forbidden,  the boundary conditions for large $R$ are different for these two types of channels. Zero inward initialization is still valid for the closed channel $\varphi_1(R)$, whereas Siegert-type outgoing boundary conditions should be adopted for the open one $\varphi_2(R)$ \cite{siegert}.  High accuracy can be achieved by using complex rotation of the coordinate $R$ \cite{moiseyev} which actually plays the role of an absorbing potential. The most important observation is that the use of the complex coordinate brings the outgoing asymptotic behavior to regularity (zero inward initialization) \cite{atabek03}. Such imposed boundary conditions are at the origin of quantization conditions leading to discrete complex resonance energies. 
 
 In practice, a plausible laser wavelength is guessed from the semiclassical coincidence criterion involving the vibrational level $\tilde{v}$ of the quasi-diabatic potential $V_1(R)$  at low intensity Eq.(\ref{eq:semiclassical2}) and one of the vibrational levels $\tilde{v}^+$ of the upper adiabatic potential $V_+(R)$ Eq.(\ref{eq:semiclassical1}). This is convenient for the search of a ZWR originating from $\tilde{v}$, which actually is nothing but $v$ at the zero field intensity limit. An important point to consider is that, due to the possibility to bring into coincidence several vibrational levels of the upper adiabatic potential $\tilde{v}^+$  with a given $\tilde{v}$, there are several wavelengths which may be convenient for a ZWR search. Such a wavelength  being fixed for a given field-free vibrational level $v$, the field strength $I$ is progressively increased. For each discrete value of $I$, starting from an initial guess for the energy coming from the previously considered value of $I$, the propagation-matching procedure leads to an accurate eigenenergy 
 $E_v=\Re(E_v)-i\Gamma_v/2$ 
 when properly converged. All calculations are conducted within the single photon absorption approximation frame of Eqs.(\ref{eq:closecoupled}), which is valid for low enough intensities, neglecting the mixing of neighboring Floquet blocks occurring within the period $2\omega$ \cite{atabek03}. 
  A typical behavior of the resonance width $\Gamma_v$ as a function of the increasing field strength $I$ displays a linear increase, and a saturation, followed by a natural decrease of the decay rates which is interpreted in terms of spatial non-adiabatic effects \cite{chrysos}. But more interesting is the observation of very fast and local collapse of the width, for specific intensities. One can finely adjust the intensity, such as to obtain resonance widths close to zero within several figures of accuracy. This shows that, at least in the single Floquet block approximation of Eqs.(\ref{eq:closecoupled}), ZWRs do exist as also predicted by the semiclassical theory. Introducing multiphoton processes will offer additional dissociation channels from which the outgoing flux could be evacuated. 
This may give rise to (small but still non zero) partial widths contributing to $\Gamma_v$ \cite{chrysos} which may 
no more strictly collapse to zero.


\subsubsection{Global methods} \label{globalmethods}

Inserting the Floquet ansatz Eq.~\eqref{eq:anzats} into the Schr\"odinger equation \eqref{TDSE} leads to an eigenvalue problem for the Floquet states $\chi_{v}$ and the associated quasienergies $E_v$,
\begin{equation}
\left[ H(t) - i\hbar \frac{\partial}{\partial t}  \right]  
\left[
\begin{array}{c}
\chi_{v,1} (R,t) \\ 
\chi_{v,2} (R,t)
\end{array} 
\right]
= E_v  
\left[
\begin{array}{c}
\chi_{v,1} (R,t) \\ 
\chi_{v,2} (R,t)
\end{array} 
\right]
,
\label{floquetev1}
\end{equation}
where $H(t)$ is the Hamiltonian introduced in Eq.~\eqref{TDSE}. 
This eigenvalue problem can be solved using a global iterative method. 
The operator $K=H(t)-i\hbar\frac{\partial}{\partial t}$ appearing in the left-hand side of Eq.~\eqref{floquetev1} is called the Floquet Hamiltonian and can be represented within a direct product basis set made of two Fourier basis sets, one for the radial coordinate and another for the time coordinate (as an additional quantum coordinate). 
We typically use 4 to 8 Floquet blocks with both open channels (describing  multi photon absorption) and closed ones (describing photon emission). With the moderate intensities used hereafter, the  Floquet block describing a single photon absorption remains the dominant one. It is to be noticed that grid method calculations can also be extended to perform  calculations including a larger number of Floquet blocks and the convergence with respect to the number of channels has been checked independently with both numerical approaches. 
In any case, if the intensity is small enough, the difference between two-channel calculations and calculations using a larger number of Floquet blocks remains very small.

Using the global representation of the Floquet Hamiltonian described above, we proceed as follows: 
For each value of the field parameters $\{I,\lambda\}$, one particular solution of Eq.~(\ref{floquetev1}) is obtained by using an initial guess for the eigenvector and modifying it by using an iterative method. We have taken advantage of the wave operator method which can handle dissipative problems and is well-adapted for searching small numbers of complex eigenvalues of a large matrix \cite{Jol3}. 
The main idea is to define an active space of interest chosen so as to possess a strong overlap with the expected unknown eigenvector.  Then we find the wave operator $\Omega $ which transforms the Floquet Hamiltonian matrix $K$ into a smaller, effective matrix within the active space,
\begin{equation}
H_{\text{eff}} = P_o [ K \Omega ] P_o,
\label{waveop}
\end{equation}
the eigenvalues of which are exact eigenvalues of the initial problem. $P_o$ is the projector on this active space
and $\Omega$ is a non-invertible matrix with a special structure, such that all columns corresponding to states out of the active space be zero. 
The wave operator can be found by effectively solving the non-linear Bloch equation \cite{killingbeck2003},
\begin{equation}
K \Omega = \Omega K \Omega.
\label{eq:bloch}
\end{equation}
The above Eq.~\eqref{eq:bloch}, together with the special structure imposed to $\Omega$, completely define the wave operator able to reduce the eigenvalue problem to an effective, small-dimensional problem.

Eq.~\eqref{eq:bloch} can be solved in an iterative way. We have used a variant of the Recursive Distorded Wave Approximation algorithm \cite{Jol3}. 
Here, the active space can be chosen as one-dimensional (the space spanned by the vibrational state from which the resonance is formed). This choice is simple and sufficient to attain convergence. This means that we calculate one resonance at a time. 
The same approach has been previously used in a slightly different context within the constrained adiabatic trajectory method (CATM) and more methodological details about the iterative equations can be found in ref. \cite{Lecl3}. 
It should be noted that the wave-operator equation (\ref{waveop}) can give either solutions to the TDSE (if suitable time-dependent absorbing potentials are included in the model to constrain the initial conditions, as in ref. \cite{Lecl3}), 
or quasienergies and the corresponding Floquet eigenstates defined in Eq. (\ref{floquetev1}) (if there are no constraints on the initial conditions, as is the case here).

In comparison with grid methods described in section \ref{gridmethods}, there is no need for an initial energy guess but we do need an initial guess for the eigenvector. At low intensities, the algorithm converges well by using a field-free vibrational state, constant over the time period, as an initial guess. Then, when the laser parameters are progressively varied it is possible to use the result obtained in the previous calculation (with slightly different parameters) as the initial guess for calculating the next eigenvector corresponding to the next laser parameters. Each new calculation costs only a few more iterations to converge toward the new quasienergy (only one iteration is often sufficient). This is an advantage of using a global iterative method, the exploration of the parameter plane is made easier. 
Another difference with grid methods is that the calculations with the global algorithm have been done using a radial complex absorbing potential to discretize the continuum, instead of using a complex rotation of the coordinate. This different implementation has very little consequence on the results, since the absorbing potential is acting over a large domain. 

The optimal ZWR paths in the $\{I,\lambda\}$ parameter plane are obtained following a two-step strategy. In a first step, we fix a low intensity and we study exhaustively the variations of the width $\Gamma_v$ with respect to the wavelength, for different resonances originating from different values of $v$. This allows us to identify the wavelengths of interest. 
Then complete ZWR paths are obtained by gradually varying the laser intensity $I$ and finding for each intensity the optimal $\lambda$ corresponding to a minimum resonance width $\Gamma_v$. 
This gives a critical set of parameters $\epsilon^{\text{ZWR}}(t) \equiv \{ I^{\text{ZWR}}(t), \lambda^{\text{ZWR}}(t) \}$ which forms an almost continuous numerical path (see section \ref{zwrmaps}).



\section{Adiabatic control theory}\label{sec3}
The purpose of optimizing laser parameters $\epsilon(t) \equiv \{E(t), \omega(t)\}$ such that the survival probability of a resonance state originating in field-free conditions from a given vibrational state $v$ be maximized, while all other resonances (originating from $v'\neq v$) are decaying fast, can be conducted within the frame of the adiabatic Floquet formalism \cite{guerin}.
The methodology goes through the following steps: (A) Introduction of the adiabatic Floquet Hamiltonian acting on an extended Hilbert space together with the conditions of its equivalence with the original TDSE of the physical space; (B) Adiabatic approximation aiming at the tracking of a single resonance in the extended space; (C) Zero-Width Resonance (ZWR) strategy in the extended space; (D) Back transformation to the physical space. 

We are hereafter developing these steps.

\subsection{Floquet Hamiltonian in the extended Hilbert space}
Starting from the TDSE:
\begin{equation}
i \hbar \frac{\partial}{\partial t} | \Phi(t)\rangle=H(t)|\Phi(t)\rangle
\label{eq:TDSE}
\end{equation}
with 
\begin{equation}
H(t)=H_0-\mu E(t) \cos [\omega(t)\cdot t].
\nonumber
\end{equation}
the first goal to achieve towards adiabaticity is to fix the rapidly growing phase $\omega(t)\cdot t$ leading to non-adiabatic fast oscillations of the control field. This is precisely done through the Floquet Hamiltonian $K(\theta)$ acting on an extended Hilbert space given as a tensorial product $\mathcal{H}\otimes L^2$ of the physical Hilbert space $\mathcal{H}$ times the space $L^2(d\theta/2\pi)$ of square integrable functions of $\theta$ defined as an additional phase, varying within the interval $[0,2\pi]$ and describing a field degree of freedom. More precisely, this adiabatic Floquet Hamiltonian involves a so-called effective frequency \cite{guerin1997,guerin}:
\begin{equation}
\omega_{\text{eff}}(t)=  \frac{d}{dt}\theta
\label{eq:omegaeff}
\end{equation}
and reads:
\begin{equation}
K(\theta) = H(t)-i \hbar \omega_{\text{eff}} \frac{\partial}{\partial \theta}.
\label{eq:floquethamiltonian}
\end{equation}
The resulting time evolution equation is:
\begin{equation}
i \hbar \frac{\partial}{\partial t} |\Psi(\theta,t)\rangle=K(\theta)|\Psi(\theta,t)\rangle
\label{eq:floquetevolution}
\end{equation}

We will now examine the relationship between the two wavepackets $|\Phi(t)\rangle$ and $|\Psi(\theta,t)\rangle$, solutions of Eq.(\ref{eq:TDSE}) and Eq.(\ref{eq:floquetevolution}), respectively.  
For doing this, $|\Psi(\theta,t)\rangle$ is expanded on the complete basis set of square integrable eigenfunctions $\{|v \rangle\}$ of $H_0$  
with the straightforward possibility to extend to energy-normalized continua. 
More precisely, $\{|v\rangle\}$ designates the complete set of eigenvectors associated with the square integrable vibrational eigenfunctions, augmented by the enumerable set of the continuum eigenfunctions sub-space. The remaining part of the expansion concerns the basis set 
$\{\langle\theta|n\rangle\}_n=\{e^{in\theta}\}_n$ of $L^2$, with time-dependent unknown coefficients $c_{v,n}(t)$:
\begin{equation}
|\Psi(\theta,t)\rangle=\sum_v \sum_n c_{v,n}(t)  \langle\theta | n \rangle |v\rangle
\label{eq:Psi}
\end{equation}
Recasting Eq.(\ref{eq:Psi}) into Eq.(\ref{eq:floquetevolution}) results in:
\begin{eqnarray}
&i \hbar \sum _{v,n} \big( \dot c_{v,n}(t) \langle\theta | n\rangle+ c_{v,n}(t) \omega_{\text{eff}}(t)  \frac{d}{d\theta}\langle\theta | n\rangle\big)|v\rangle& \nonumber \\
&= H(t)\sum_{v,n} c_{v,n}(t) \langle\theta | n\rangle|v\rangle& 
\label{eq:cvn}
\end{eqnarray}
The notation $\dot c$ is for the time-derivative of $c$.
It is important to notice that Eq.(\ref{eq:cvn}) holds for any time $t$ and for any phase $\theta$, taken up to here as two independent variables. 
More specifically, from now on $\theta$ is taken as:
\begin{equation}
\theta=\omega(t) \cdot t
\label{eq:theta}
\end{equation}
This choice being done for $\theta$, Eq.(\ref{eq:cvn}) reads:
\begin{equation}
i \hbar \frac{\partial}{\partial t} \sum_{v,n}  c_{v,n} \langle\theta | n \rangle |v\rangle= H(t) \sum_{v,n} c_{v,n} \langle\theta | n \rangle |v\rangle
\end{equation}
or equivalently:
\begin{equation}
i \hbar \frac{\partial}{\partial t}| \Psi(\theta,t)\rangle=H(t)|\Psi(\theta,t)\rangle
\end{equation}
which shows that if $|\Psi(\theta,t)\rangle$ is a solution of Eq.(\ref{eq:floquetevolution}), then $|\Phi(t)\rangle = |\Psi(\theta, t)\rangle$ is in turn a solution of Eq.(\ref{eq:TDSE}), with $\theta$ defined by Eq.(\ref{eq:theta}).
The fast oscillating behavior of $ \vert \Phi(t)\rangle$ is now recast in $\langle\theta | n \rangle$, the Fourier components $c_{v,n}(t)$ of $\Psi(\theta,t)\rangle$ being henceforth slowly varying functions of time.

As a conclusion, the two control strategies aiming at a vibrationally selective survival through an adiabatic tracking of a ZWR, namely: (i) finding the laser parameters $\epsilon(t)\equiv\{E(t), \omega(t)\}$ such that $|\Phi(t)\rangle$, solution of Eq.(\ref{eq:TDSE}) follows this ZWR, or, (ii) finding the laser parameters $\epsilon(t)\equiv\{E(t), \omega_{\text{eff}}(t)\}$ such that $|\Psi(\theta,t)\rangle$, solution of Eq.(\ref{eq:floquetevolution}) follows the ZWR, are formally identical, with the choice of Eq.(\ref{eq:theta}) for $\theta$. In the following it is the adiabatic strategy (ii) which is retained.

\subsection{Adiabatic approximation}
We are now looking for an adiabatic evolution such that at all times $t$, the solution $|\Psi(\theta,t)\rangle$ follows a specific single resonance eigenvector $|\chi_v\rangle$ of the adiabatic Floquet Hamiltonian labeled by its corresponding field-free parent state $|v \rangle$:
\begin{equation}
K(\theta)|\chi_v\rangle = E_v |\chi_v\rangle. 
\label{floquetev2}
\end{equation}
This equation should be understood in the following way: At each time $t$, the field parameters $\epsilon(t)\equiv\{E(t), \omega_{\text{eff}}(t)\}$ give rise to a Hamiltonian $K(\theta;\epsilon(t))$ with resonance eigenfunctions $\langle \theta |\chi_v\rangle = \chi_v\{\theta; \epsilon(t)\}$ and eigenvalues $E_v\{\epsilon(t)\}$.
The adiabatic approximation for $|\Psi(\theta,t)\rangle$ is nothing but:
\begin{equation}
|\Psi_v^{ad}(\theta, t)\rangle = \exp{\big [-i/\hbar \int_0^t E_v\{\epsilon(t')\}dt'}\big]~ |\chi_v\{\theta, \epsilon(t)\}\rangle.
\label{eq:psiad}
\end{equation}
We note, for completeness, that the exact wavefunction $|\Psi(\theta,t)\rangle$ could be given as an expansion, on the basis set of the resonance eigenvectors $\{|\chi_{v'}\rangle\}_{v'}$ or their analogs $\{| \Psi_{v'}^{ad}\rangle\}_{v'}$ including the corresponding phase $\exp{\big [-i/\hbar \int_0^t E_{v'}\{\epsilon(t')\}dt'}\big]$:
\begin{equation}
|\Psi(\theta,t)\rangle= \sum_{v'} d_{v'}(t) |\Psi_{v'}^{ad}(\theta, t)\rangle 
\end{equation}
The initial condition describing the system in its vibrational state $v$, implies:
\begin{equation}
d_v(0)=1~ ; ~d_{v'}(0) = 0 ~ ; ~\forall v' \neq v
\end{equation}
with the consequence :
\begin{equation}
|\Psi(0,0)\rangle = |\Phi(0)\rangle = |\chi_v(0)\rangle
\end{equation}
which actually is the initial field-free vibrational wavefunction.

\subsection{Zero-Width Resonance strategy}

The full control strategy is built in two steps: (i) Trapping the system into a single eigenvector $|\chi_v\rangle$ of the adiabatic Floquet Hamiltonian; and (ii) Designing a pulse with field parameters such that this eigenstate presents the lowest (zero, if possible) dissociation rate. The requirement of step (i) leading to the approximation of Eq.(\ref{eq:psiad}), is expected to be reached for slowly enough (adiabatically) varying parameters $\epsilon(t)$ avoiding any degeneracy between complex eigenvalues $E_{v'}\{\epsilon(t)\}$, at all times~$t$~\cite{nimrod}.
The second step (ii) involves an optimal choice for the field parameters:
\begin{equation}
\epsilon^*(t)\equiv\{E^*(t), \omega_{\text{eff}}^*(t)\}
\nonumber 
\end{equation}
such that:
\begin{equation}
\Im[E_v\{\epsilon^*(t)\}]=0 ~~ \forall t
\label{eq:imag}
\end{equation}
$\Im(E_v)$ being the imaginary part of the energies of these field-induced resonances.  
We note that Eq.(\ref{eq:imag}) stands precisely for the search of a ZWR path (originating from $|v \rangle$) in the amplitude/frequency parameter plane (or equivalently, intensity $I$ / wavelength $\lambda$) as a function of $t$. In other words, the star * in Eq.(\ref{eq:imag}) could be replaced by the superscript ZWR, such that:
\begin{equation}
\epsilon^{\text{ZWR}}(t)\equiv
\left\{
\begin{array}{c}
E^{\text{ZWR}}(t) = [I^{\text{ZWR}}(t)]^{1/2}, \\ 
\omega_{\text{eff}}^{\text{ZWR}}(t) = 2\pi c/\lambda^{\text{ZWR}}(t) 
\nonumber 
\end{array}
\right\}
\end{equation}
be the control field of the evolution monitored by  the adiabatic Floquet Hamiltonian of the extended Hilbert space Eq.(\ref{eq:floquetevolution}). 
Following such a resonance associated with a quasi-infinite lifetime is very favorable for obtaining a good adiabatic approximation in this dissipative context. 
Following the less dissipative eigenstate is actually a requirement of the adiabatic theorem in presence of dissipation \cite{nenciu}. 
In any other situation, there is a risk of loosing population in the selected resonance and to magnify non-adiabatic transitions due to this relative population loss with respect to other eigenstates. 
Following a ZWR automatically avoids this magnification of non-adiabatic contaminations, which could occur if the adiabatic approximation is done on a resonance associated with a non-zero decay rate. 

On mathematical grounds, it is also to be noticed that fulfilling the requirement of Eq.(\ref{eq:imag}) leads to the possibility of solving the challenging issue of adiabatic transport involving passages through continuum spectra. In other words, ZWRs are good candidates for a full adiabatic Floquet treatment as initially derived for pure bound states.

\subsection{Back transformation to the physical Hilbert space}
The purpose is now to obtain the optimal field parameters acting in the original Hilbert space where the evolution is monitored by the TDSE displayed in Eq.(\ref{eq:TDSE}). This is achieved by solving the first order differential equation Eq.(\ref{eq:omegaeff}), with $\omega^{*}(t)$ as the unknown function:
\begin{equation}
\frac{d}{dt}\omega^*(t) \cdot t+\omega^*(t)-\omega_{\text{eff}}^{\text{ZWR}}(t)=0
\label{eq:domega}
\end{equation}
Here again the star * stands for the optimal control solution. A particular solution of Eq.(\ref{eq:domega}) is obtained as:
\begin{equation}
\omega^*(t)= \frac{1}{t} \int_0^t \omega_{\text{eff}}^{\text{ZWR}}(t') dt'
\end{equation}
which satisfies the expected initial condition, i.e.:
\begin{equation}
\lim_{t \rightarrow 0}  \frac{1}{t} \int_0^t \omega_{\text{eff}}^{\text{ZWR}}(t') dt'=\omega_{\text{eff}}^{\text{ZWR}}(0)=\omega^*(0)
\end{equation}
Finally, the external laser field which fulfills the requirements of the control objective, in the physical Hilbert space is:
\begin{eqnarray}
\mathcal{E}^*(t) &=& E^*(t) \cos{[\omega^*(t) \cdot t]}  \nonumber \\
&=&  E^{\text{ZWR}}(t) \cos \left({\int_0^t \omega_{\text{eff}}^{\text{ZWR}}(t') dt'} \right)
\label{adiab_pulse1}
\end{eqnarray}
which can be written in terms of intensity/wavelength parameters as:
\begin{equation}
\mathcal{E}^*(t)= [I^{\text{ZWR}}(t)]^{1/2} \cdot  \cos \left({\int_0^t 2\pi c/\lambda^{\text{ZWR}}(t') dt'} \right)
\label{adiab_pulse}
\end{equation}


\subsection{Numerical methods and test of the adiabatic character \label{nummethod} }

All the dynamical results presented below have been obtained by rigorously solving the TDSE.
We have used two independent methods to validate the numerical results.

We have first used the split-operator scheme~\cite{feit} which is based on a differential propagation over small time steps. 
Split-operator calculations have been done either with a grid Fourier DVR basis set associated with the internuclear distance coordinate or within the bare vibrational eigenbasis, the splitting being done between kinetic and potential operators or between the diagonal and dipolar coupling terms, respectively.

We have also confirmed all the results by using a recently developed iterative algorithm 
based on the time-dependent wave operator formalism in which Fourier decomposition is also used 
to describe the time coordinate \cite{Lecl2015,Jol2016}, allowing a much smaller number of discretization points. In this framework, the time-dependent wave operator $\Omega(t)$ is calculated and used to deduce the true dynamics from the dynamics within a smaller dimensional subspace which includes only a small part of the bare vibrational (bound and possibly continuum) states.
The evolution operator issuing from the small-dimensional subspace is calculated as
\begin{equation}
U(t,0;H)P_o=\Omega (t) U(t,0;H_{\text{eff}}).
\label{evop}
\end{equation}
where $H_{\text{eff}}(t)=P_o H(t)\Omega(t)$ is a time-dependent effective Hamiltonian governing the dynamics within the active subspace of projector $P_o$. 
This effective Hamiltonian is similar to the one of Eq.(\ref{waveop}) but is now defined over the entire duration of the chirped pulse, while Eq.(\ref{waveop}) is valid over a single optical period, at fixed frequency and intensity. 

We shall point out that this concept of effective Hamiltonian is not only a computational intermediate, it also naturally gives an indication about the adiabatic or non-adiabatic character of the dynamics during the interaction. 
If the active subspace is one-dimensional (the space spanned by the vibrational state $\vert v \rangle $ associated with 
the selected ZWR $\vert \chi_v \rangle $), then the effective Hamiltonian leads to an effective energy \cite{jolicard2003},
\begin{equation}
E^{v}_{\text{eff}} (t ) = \langle v \vert H_{\text{eff}} \vert v \rangle = \frac{\langle v \vert H(t) \ket{\Phi (t)}}{\langle v \ket{\Phi(t)}},
\label{eeff1}
\end{equation}
where $\ket{\Phi(t)}$ is a solution of the TDSE (\ref{eq:TDSE}). 
This quantity follows some trajectory in the complex plane which illustrate the more or less
adiabatic character of the dynamics. 
Assuming a perfect adiabatic dynamics and using Eq. (\ref{eq:psiad}), 
the effective energy can also be expressed as the sum of two terms   \cite{jolicard2003,Lecl2013}
\begin{equation}
E_{\text{eff}}^{v,\text{ad}} (t)
= E_v \{ \epsilon(t) \} 
+ i \hbar \frac{ \langle v \vert \frac{\partial}{\partial t} \ket{\chi_v\{ \theta, \epsilon(t) \} } }{ \langle v \ket{ \chi_v \{ \theta, \epsilon(t) \} } },
\label{eeff2}
\end{equation}
where $\ket{ \chi_v \{ \theta, \epsilon(t) \} }$ and $E_v \{ \epsilon(t) \}$ are the selected instantaneous Floquet state and eigenvalue, corresponding to the field parameters $\epsilon(t)$ (cf. Eq. (\ref{floquetev2})). 
Following a ZWR path, the first term $E_v \{\epsilon(t) \}$ will approximately follow a straight line along the real axis during the pulse, starting from the initial field-free vibrational energy. 
The second term in Eq. (\ref{eeff2}) depends on $\ket{ \chi_v \{ \theta, \epsilon(t) \} }$ which is time-periodic with a chirped period, $T(t)=2 \pi / \omega^{\text{ZWR}}_{\text{eff}}(t)$. This term may show rapid oscillations in the complex plane. 
For a given ZWR path, two different trajectories can be calculated: $E^{v}_{\text{eff}} (t )$, using the exact wavefunction
(direct use of Eq. (\ref{eeff1}) with the solution of Eq. (\ref{eq:TDSE}) including all the possible non-adiabatic transfers), 
and $E_{\text{eff}}^{v,\text{ad}} (t)$, expected from the adiabatic approximation using Eq. (\ref{eeff2}). 
The quality of the matching between these two complex trajectories will be used in section \ref{adiatransp} as an indicator for the quality of the adiabatic following.

\section{Vibrational cooling by filtration} \label{sec4}

This Section is devoted to the potentiality of ZWRs implemented in an adiabatic control scenario to reach a filtration strategy for molecular vibrational cooling with Na$_2$ as an illustrative example. The following application deals with a realistic situation  since the formation of ultracold sodium molecules has already been demonstrated~\cite{fatemi}. The transposability to other diatomic molecules, especially to alkali dimers, requires only small changes due to the similarities of their potentials.

\subsection{The model.}
Transitionally and rotationally cold, tightly bound and vibrationally hot Na$_2$ species are experimentally produced by photoassociation in a metastable bound state $^3\Sigma_u^+ (3^2S+3^2S)$, considered as a ground state (referred to as state u). Typically, vibrational levels with quantum numbers $v \geq 8$ are prepared. The laser controlled filtration strategy consists in applying an electromagnetic field with wavelengths around 570$nm$ which couples state u with a repulsive, thus dissociating excited (1)$^3\Pi_g (3^2S+3^2P)$ electronic state (referred to as state g). The corresponding Born-Oppenheimer potential energy curves $V_{1,2}(R)$ and the electronic transition dipole moment $\mu_{12}(R)$ between states g and u are taken from the literature \cite{magnier,aymar, AtabekPRL} and references therein. Our specific model refers to a rotationless field-aligned molecule in a single spatial dimension (namely, the internuclear distance $R$). Such a frozen rotation approximation is validated when comparing the short pulse durations in consideration (12$ps$) with 
the long rotational periods of Na$_2$ (estimated as hundreds of $ps$). Finally, Na$_2$ reduced mass which is involved in the kinetic energy operator $T_N$ of Eq.(2) is taken as 20963.2195$au$.

\subsection{ZWR maps in the laser parameter plane.} \label{zwrmaps}
Solving time-independent coupled equations Eq.(\ref{eq:closecoupled}) with Siegert boundary conditions for a set of continuous wave cw laser parameters $\{E,\omega\}$ or equivalently $\{I,\lambda\}$, gives rise to resonances with complex eigenvalues $E_v$ correlating, in field-free conditions, with the real vibrational eigenenergies $E_v$. We are actually interested in finding specific couples of field parameters for which the imaginary part of resonance eigenvalues are close to zero.

\begin{figure*}
	\includegraphics[width=0.5\linewidth]{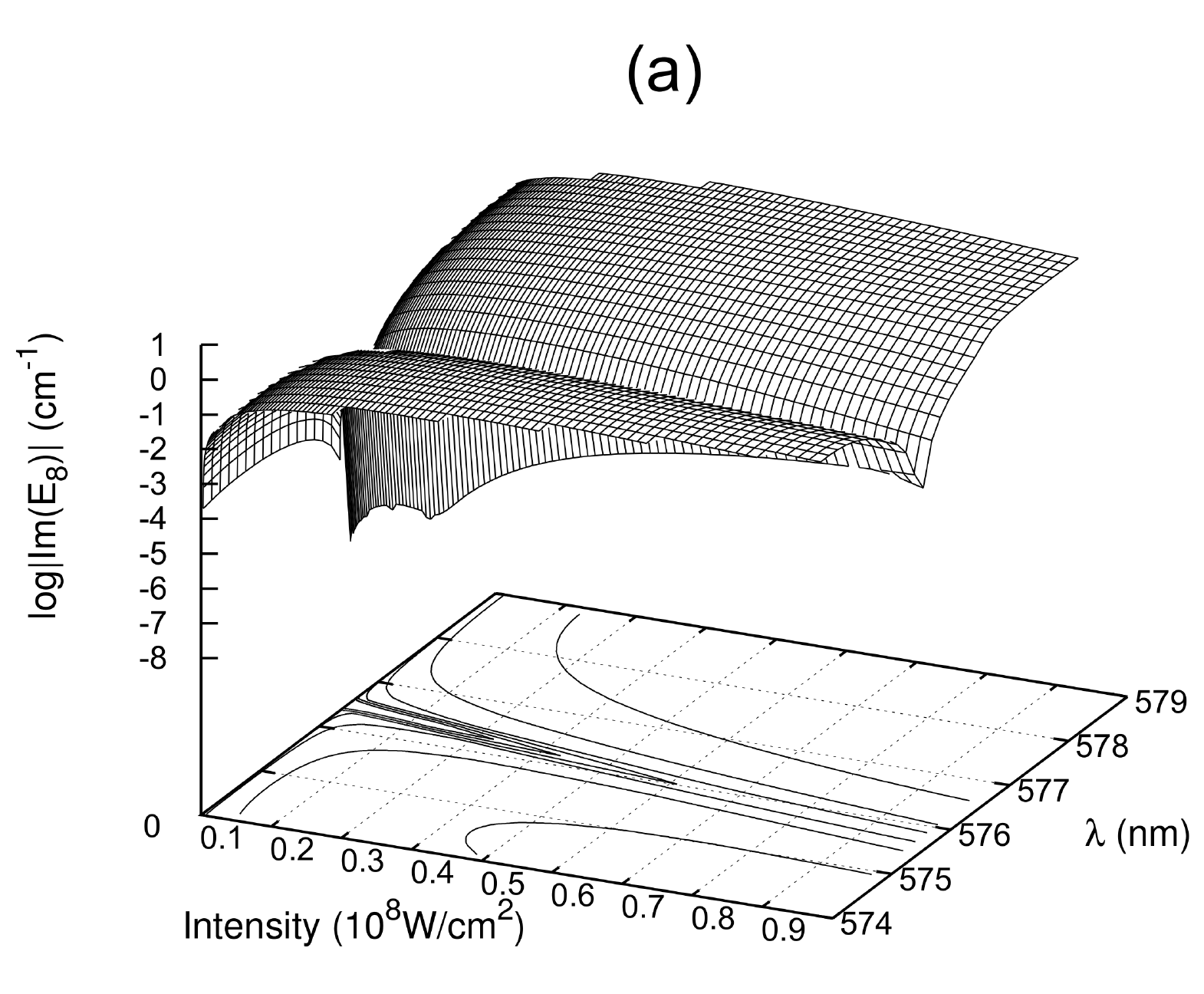}
	\includegraphics[width=0.4\linewidth]{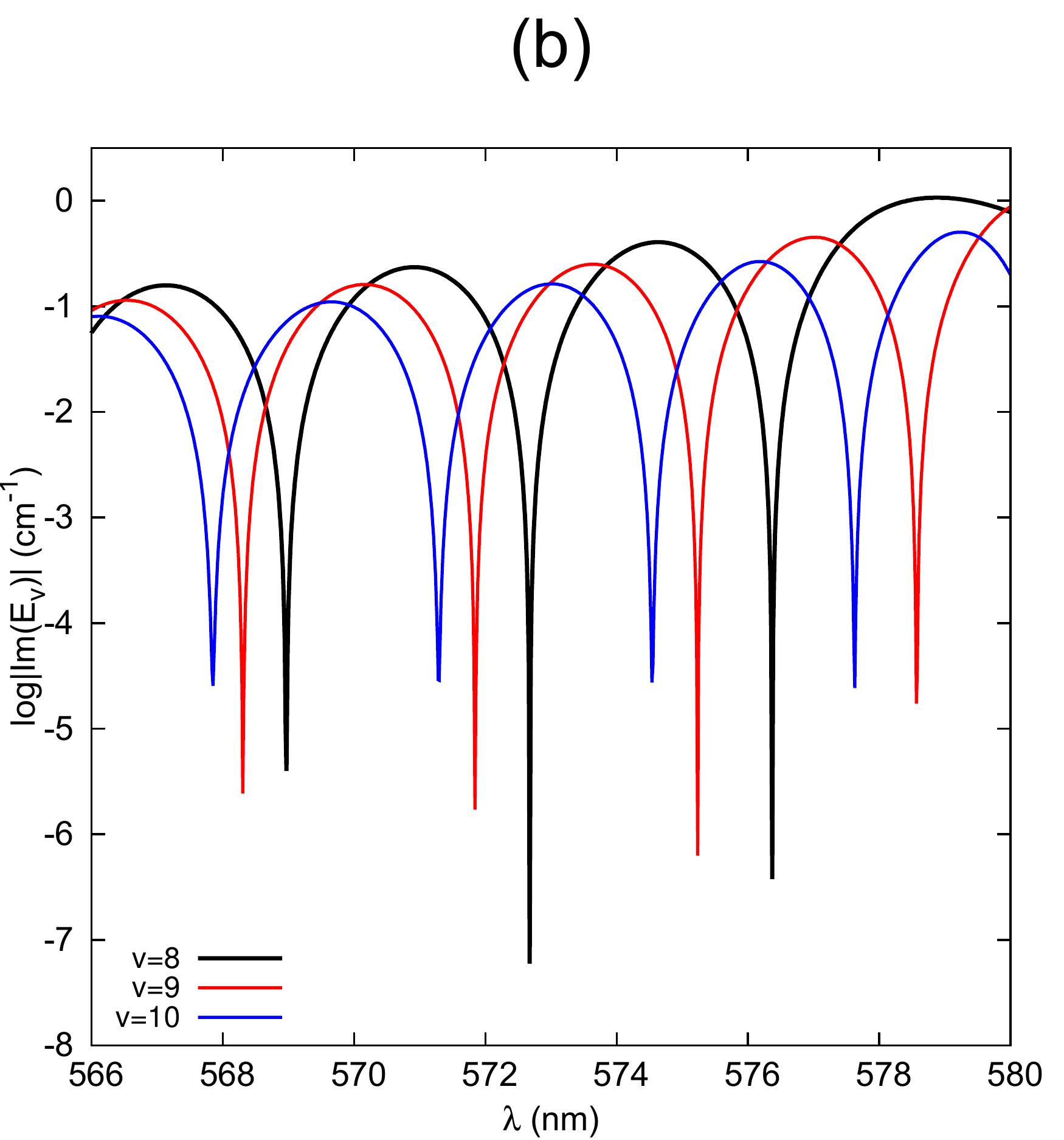}
	\caption{Morphology of ZWRs originating from state $v=8$ of Na$_2$. 
		Left frame (a): $\Im (E_8)$ as a function of the laser wavelength $\lambda$ and intensity (logarithmic scale).
		Right frame (b): $\Im (E_v)$ as a function of the wavelength $\lambda$ for resonances originating from $v=8, 9, 10$ for a laser intensity $I=0.14 \times 10^8$ W.cm$^{-2}$ (logarithmic scale).}
	\label{morphology_zwr}
\end{figure*}
The general morphology of ZWRs is illustrated in Fig.(\ref{morphology_zwr}) which displays an exploratory calculation in the $\{I,\lambda\}$ parameter plane. The left panel shows the imaginary part of $E_8$ (in logarithmic scale) as a function of laser intensity and wavelength covering a 5$nm$ wide window in the vicinity of $\lambda=576nm$. The right panel displays the imaginary parts of $E_v, (v=8,9,10)$ as a function of the laser wavelength for a fixed intensity $I=0.14 \times 10^8$ W.cm$^{-2}$. Within numerical inaccuracies inherent to the evaluation of such very small width resonances (typically less than $10^{-6} cm^{-1}$), we observe that there are several couples of critical wavelengths and intensities producing ZWRs originating from a single vibrational level $v$.

Referring to the semiclassical analysis, the quantum destructive interference pattern from phase (or energy) coincidences in Eq.(\ref{eq:Gammasem}) leads to the choice of photon frequencies picking out the couple of electronic states which take part in the process by appropriately dressing the corresponding adiabatic potentials. The field intensity is the remaining parameter which allows an accurate phase coincidence, aiming at reaching ZWRs. Due to the relatively moderate intensities we are referring to, a single photon model with only two electronic states offers enough accuracy for converged calculations, obtained using the grid and global methods presented in section \ref{computmethods}.
Clearly, for a given field-free vibrational state, the destructive interference leading to a ZWR cannot take place, whatever the intensity be, if an avoided curve crossing position exceeds the right turning point of the vibrational level in consideration. In particular, for $v=8$ this fixes a maximum value for $\lambda$ at about 577$nm$. Guided by the semiclassical analysis, a tentative wavelength of $\lambda=576nm$ brings roughly into coincidence the $v=8$ level with the $v_+=0$ level of the upper adiabatic potential. The fine adjustment is obtained referring to the full quantum Floquet approach by tuning the remaining field parameter, i.e., the intensity $I$, such as to produce the ZWR within accuracies of typically $10^{-7}cm^{-1}$ concerning the widths.
\begin{figure}
	\includegraphics[width=0.9\linewidth]{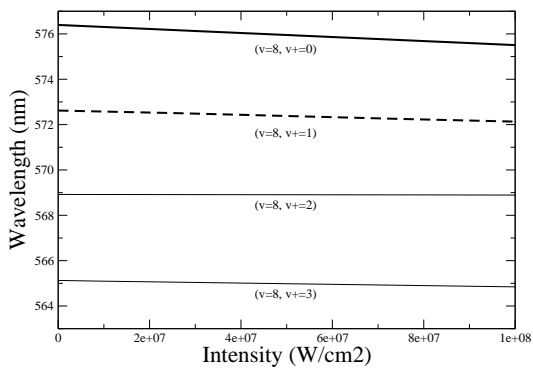}
	\caption{ZWR paths originating from $v=8$ in the laser parameter plane $\{\lambda,I\}$. The labels ($v=8$, $v_+$=0,1,2,3) refer to energy coincidences between $v=8$ on one hand, and on the other hand, the levels $v_+$ of the upper adiabatic potential of the semiclassical picture.}
	\label{loci_zwr}
\end{figure}

As has previously been observed \cite{AtabekRC}, within an intensity window not exceeding $10^8W/cm^2$, ZWRs are very close to being positioned along a straight line in the $\{\lambda,I\}$ parameter plane. Figure (\ref{loci_zwr}) displays such a behavior which seems generic for all ZWR paths. These are labeled as ($v=8,v_+$) referring to the field-free initial vibrational state $v=8$ and the successive $v_+=0,1,2,3$ levels of the upper field-dressed adiabatic potential of the semiclassical picture brought into coincidence using different wavelengths. It is worthwhile noting that the semiclassical picture is merely referred to for an initial guess of laser parameters and also for a systematic labeling of ZWRs paths. Figure (\ref{loci_zwr}) illustrates the general morphology of ZWR in the laser parameter plane. It could obviously be extended to additional paths based on coincidences involving $v=8$ with $v_+$=4,5,...

\subsection{Adiabatic transport dynamics.} \label{adiatransp}
We are now looking for the dynamical signature of a ZWR on a vibrational population transfer using the adiabatic transport strategy of Section III. We proceed by shaping a laser pulse envelope optimized for such an adiabaticity with a total duration not exceeding 12$ps$ to fulfill the frozen rotation requirement. The remaining part of the strategy consists in chirping the pulse such as to combine at each time the wavelength and intensity corresponding to the following of a ZWR path as obtained from the Floquet theory and illustrated in Figure (\ref{loci_zwr}). 
For convenience a linear fit is used for representing these paths:
\begin{equation}
 \lambda^{\text{ZWR}}(nm)  = a \, I^{\text{ZWR}} (10^8W/cm^2) + b.
 \label{eq:linear}
\end{equation}
The coefficients $a$ and $b$ are separately calculated for each of the different paths ($v=8,v_+$). Taking  ($v=8,v_+=0$) as
 an illustrative example we get 
$a=-0.7723229$ and $b=576.3668$. 
 The additional parameters involved in the pulse envelope could be obtained by optimal control algorithms aiming at the minimization of the population decay due to non-adiabatic contamination in the transport. We simplified the procedure and based on our previous experience, after a few attempts, we have used linear ramps of intensity,
\begin{equation}
I(t) = \left\{ 
\begin{array}{ll}
\frac{I_{\text{max}}}{T_{1/2}} \cdot t & \text{ for } t \in [0,T_{1/2}] \\ 
I_{\text{max}} \left( 2 - \frac{t}{T_{1/2}} \right) & \text{ for } t \in [T_{1/2}, T_{\text{tot}}] 
\end{array} 
\right.
\label{eq:enveloppe}
\end{equation}
where $T_{1/2} = T_{\text{tot}} /2$ is half the total pulse duration. 
The wavelenght is continuously chirped according to Eq.~(\ref{eq:linear}) to follow the ZWR path. 
\begin{figure}
	\includegraphics[width=0.9\linewidth]{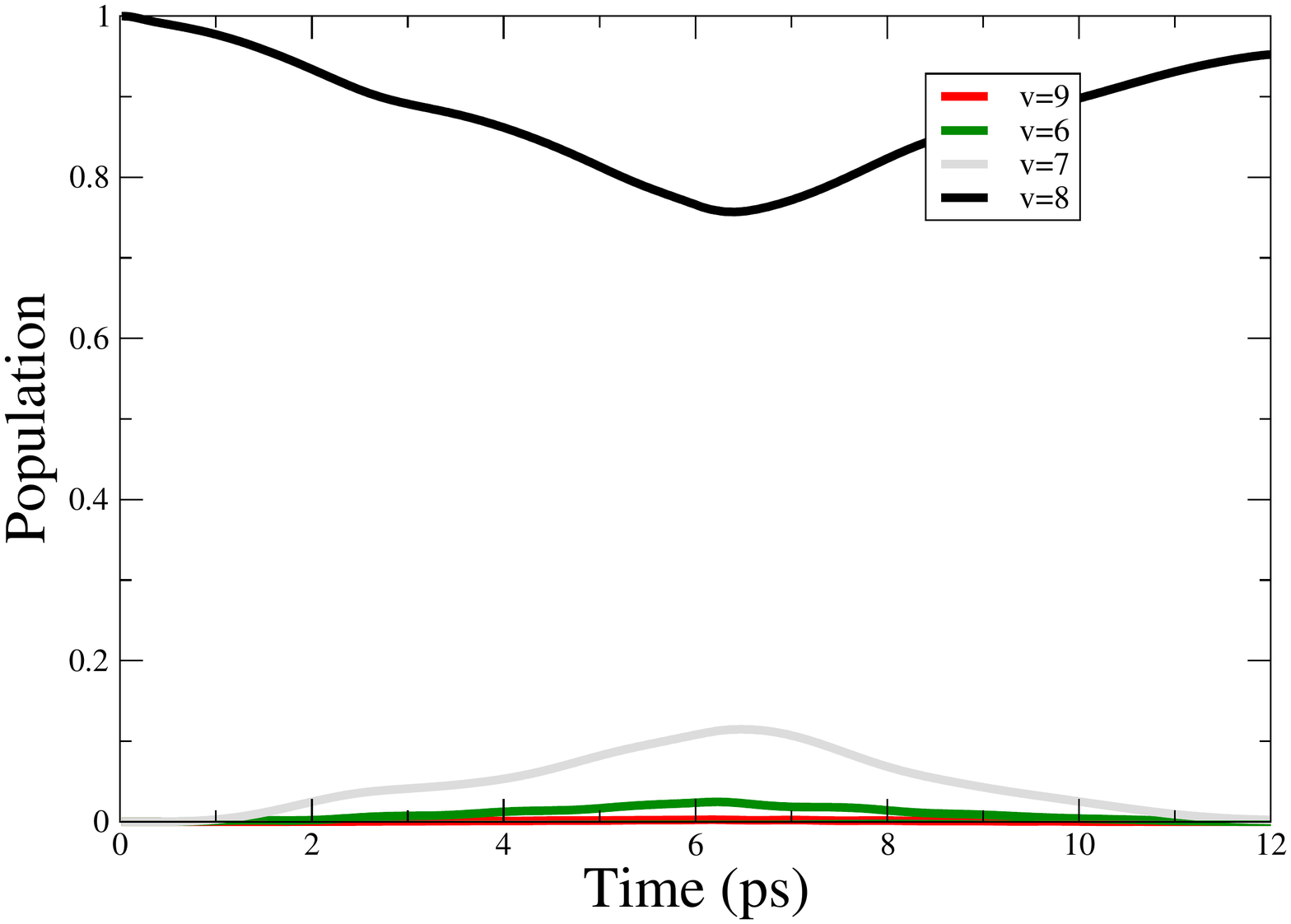}

 	\vspace{-0.4cm}
	\includegraphics[width=0.9\linewidth]{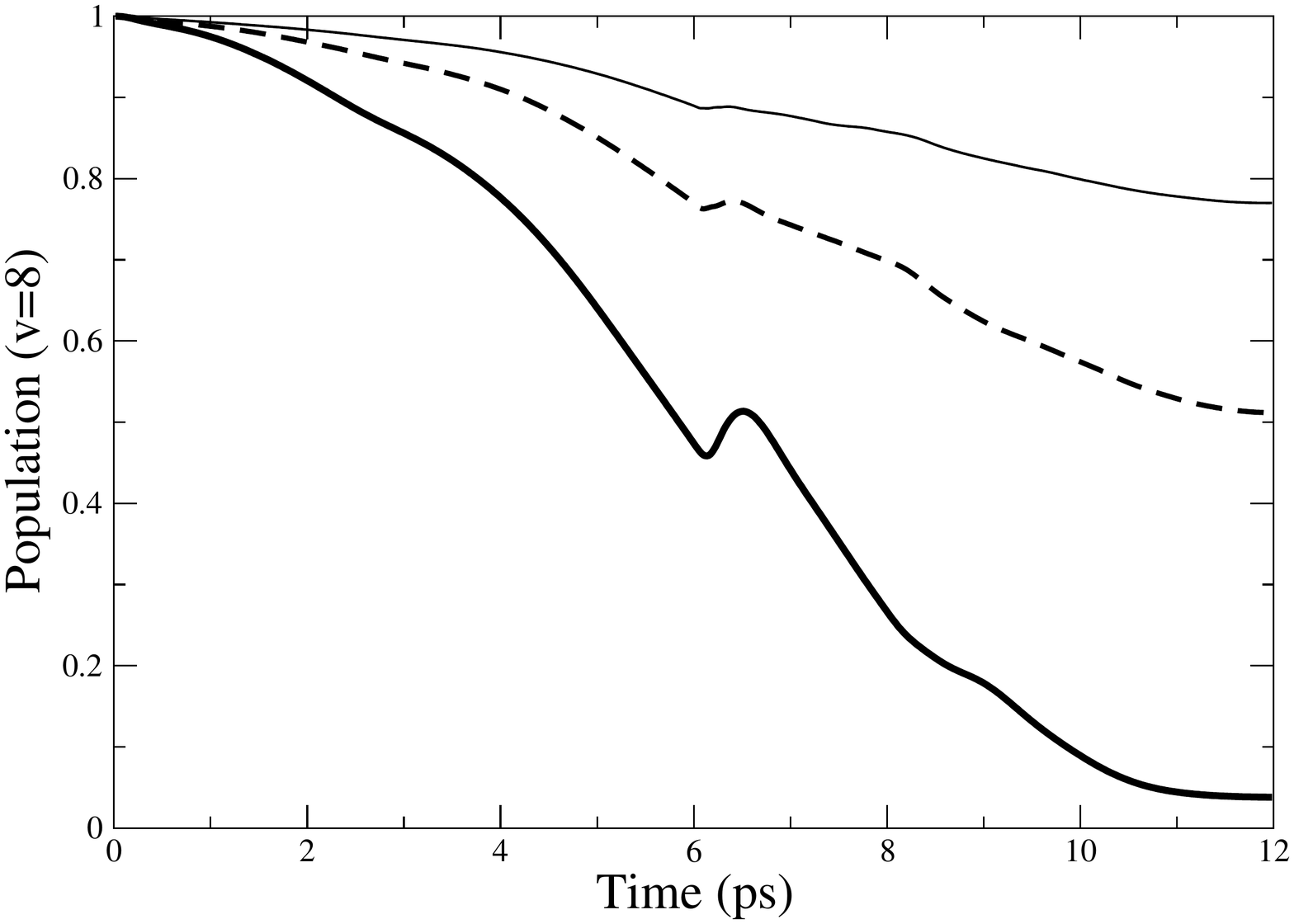}
	\caption{Vibrational populations of Na$_2$ as a function of time. 
		(a) For the adiabatic transport based on a laser pulse resulting from Eqs.(\ref{eq:enveloppe}) and (\ref{adiab_pulse}), following the ZWR path ($v=8$, $v_+=0$) of Fig. (\ref{loci_zwr}): $v=8$, solid black line, $v=6,7,9$ solid green, grey and red lines respectively
		(b) For the instantaneous ZWR frequency strategy of Eq.(\ref{eq:ins_pulse}). Only the populations of $v=8$ are displayed for three ZWR paths ($v=8$, $v_+=0$) thick solid line, ($v=8$, $v_+=1$) dashed line and ($v=8$, $v_+=2$) thin solid line, following the notations of Fig. (\ref{loci_zwr}) .}
	\label{adiab_ins}
\end{figure}
Figure \ref{adiab_ins} (upper panel) illustrates our strategy with a total pulse duration of 12$ps$, long enough to produce almost complete depletion of neighboring level populations $v=6,7,9$. During the transport dynamics, the initial population of $v=8$ is partially shared with $v=7,6$ and 9 (in decreasing importance). As expected from adiabaticity, at the end of the pulse, the major part of shared populations is recovered back by $v=8$ which ends up with a final population of 95$\%$. The quality of adiabaticity which is achieved can be measured by less than 5$\%$ of the $v=$8 population escaped towards the dissociative continuum.

At that respect, it is worthwhile comparing the efficiency of the adiabatic transport control we are presenting in this work Eq.(\ref{adiab_pulse}) to the one much more naive, we have previously discussed about \cite{AtabekRC}, where the pulse was shaped referring to an instantaneous ZWR frequency strategy:
\begin{equation}
\mathcal{E}(t)= [I^{\text{ZWR}}(t)]^{1/2} \cdot  \cos [ (2\pi c/\lambda^{\text{ZWR}})t]
\label{eq:ins_pulse}
\end{equation}
Figure \ref{adiab_ins} (lower panel) displays the results for the same $v=$8 and the ZWR path identified in Eq.(\ref{eq:linear}), but with the field of Eq.(\ref{eq:ins_pulse}) with the envelope given by Eq.(\ref{eq:enveloppe}). The efficiency with respect to the protection against dissociation is completely washed out, with only 5$\%$ of the population remaining undissociated. Similar calculations done with the other ZWR paths, taken from Fig.(\ref{loci_zwr}), lead to intermediate efficiency, namely, 50$\%$ for ($v=8$, $v_+=1$) and 75$\%$ for ($v=8$, $v_+=2$), but still clearly illustrating that the instantaneous strategy is far being an adiabatic transport, as has been previously proposed \cite{AtabekRC}.

\begin{figure}
\begin{center}
\includegraphics[width=0.9\linewidth]{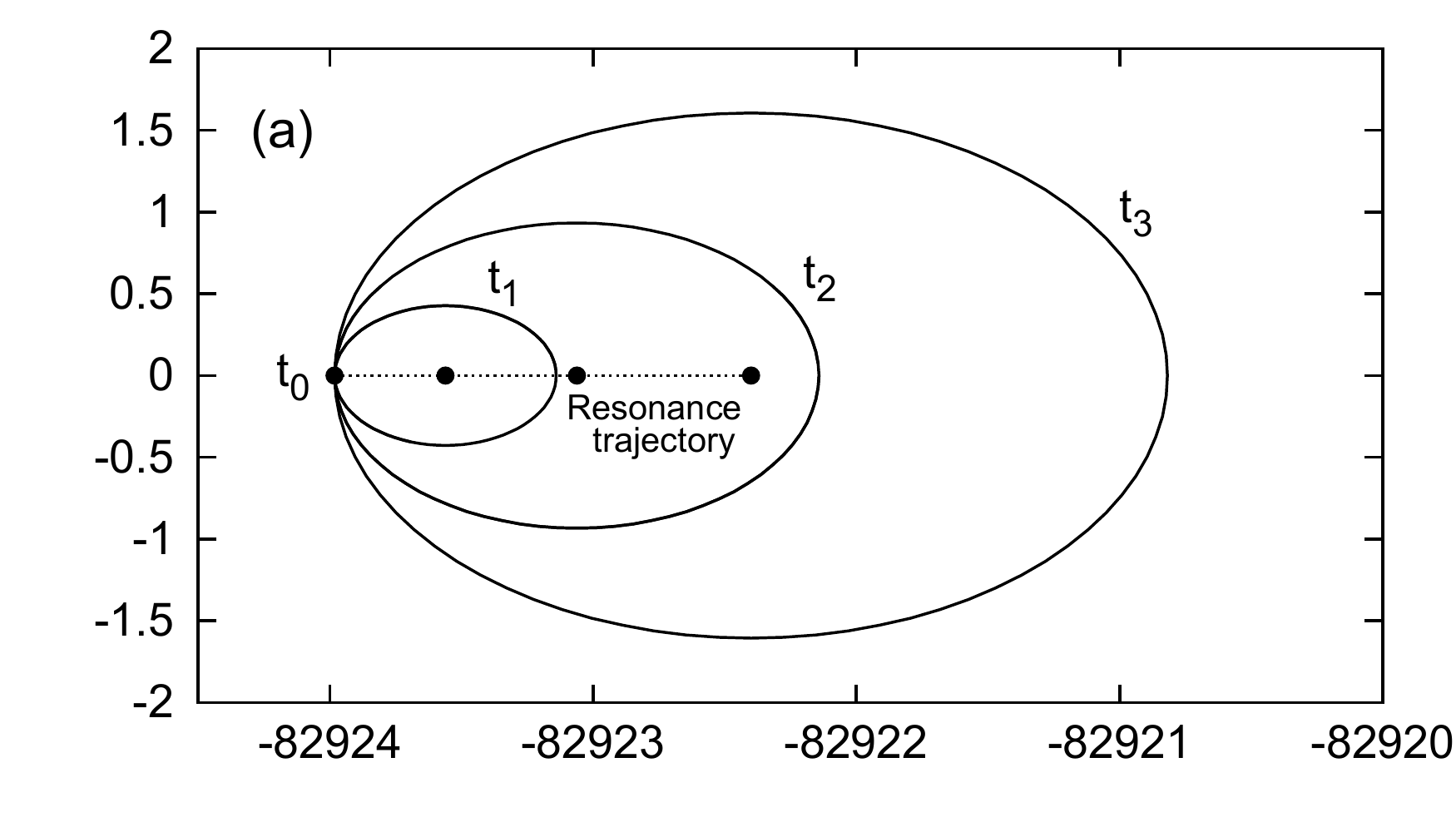}
\includegraphics[width=0.9\linewidth]{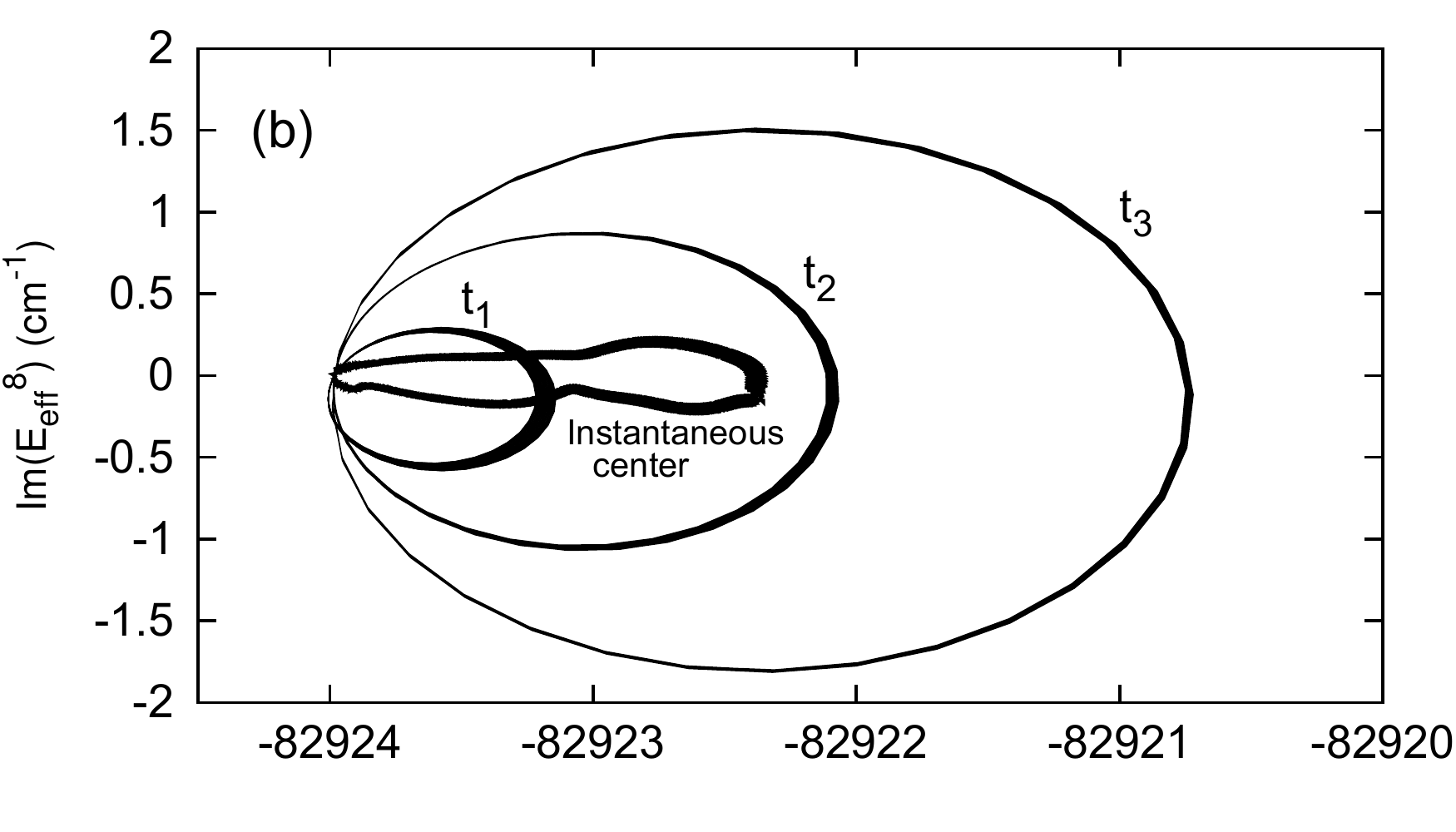}
\includegraphics[width=0.9\linewidth]{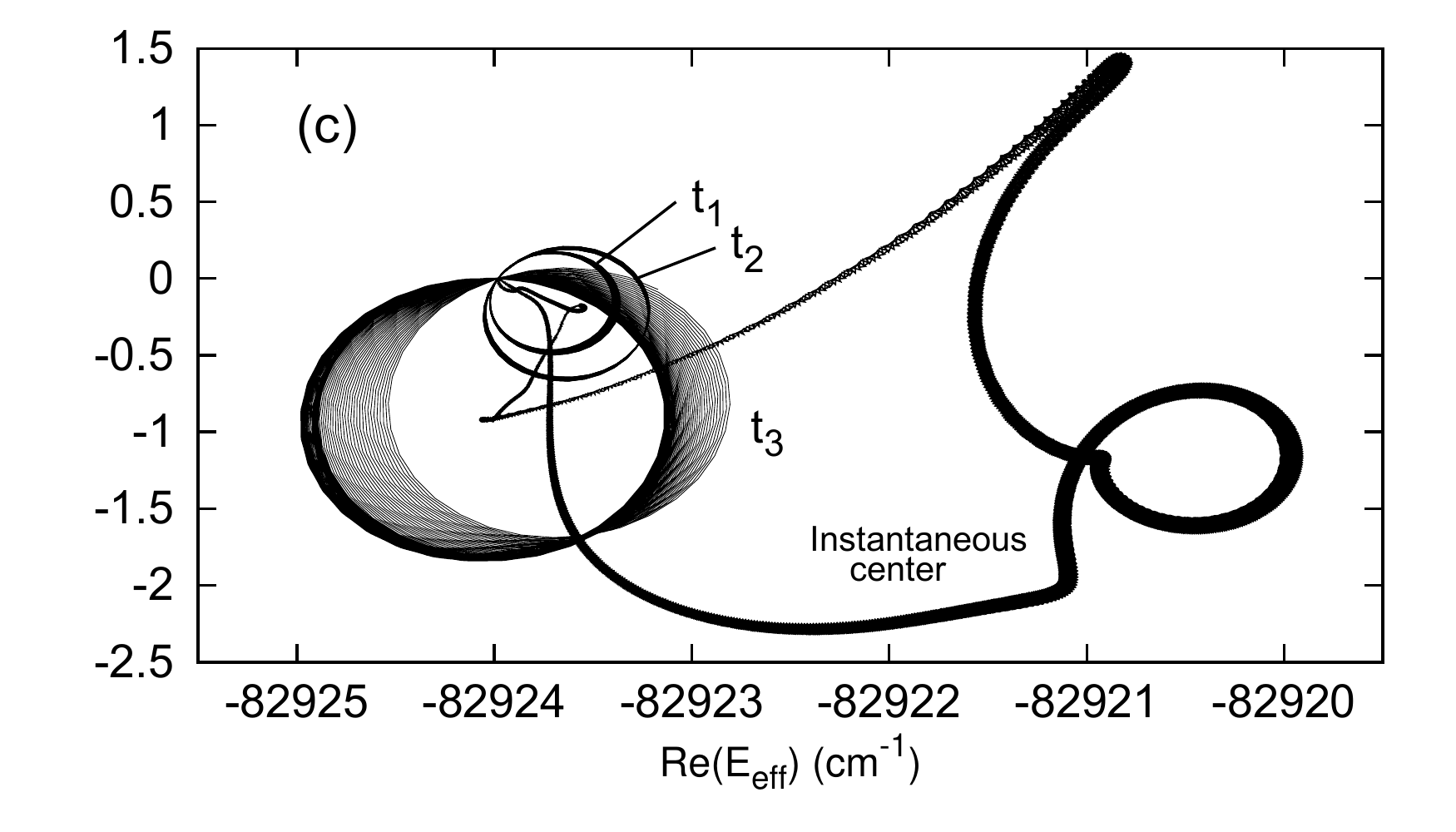}
\end{center}
\caption{Complex trajectory followed by the effective energy associated with state $v=8$ (cf. equations (\ref{eeff1}) and (\ref{eeff2})) during the dynamics following the ZWR path. 
(a) Ideal trajectories expected from the adiabatic approximation following a single Floquet eigenstate. 
(b) Results from the TDSE using the pulse of Eq. (\ref{adiab_pulse1}) and (\ref{eq:enveloppe}) with adiabatic phase $\int_0^t \omega_{\text{eff}}^{\text{ZWR}}(t') dt'$. 
(c) Results from the TDSE using the pulse of Eq. (\ref{eq:ins_pulse}) with the naive phase $\omega_{\text{eff}}^{\text{ZWR}} \cdot t$. 
The ellipses correspond to short extracts of the complex parametric curve $E_{\text{eff}}^8(t)$
around three different times corresponding to given intermediate values of the field amplitude during the rise of the pulse: $E(t_1) = 0.5 E_{\text{max}}$, $E(t_2) = 0.75 E_{\text{max}}$ and $E(t_3) = E_{\text{max}}$. For panels (b) and (c) the instantaneous center is the complex trajectory obtained by performing a rolling mean of $E_{\text{eff}}^v(t)$ over one optical cycle (see text). 
Note that the scales are slightly different in panel (c). }
\label{fig_eeff}
\end{figure}

Fig. \ref{fig_eeff} is another way of checking the adiabaticity of the dynamical process.
Following the discussion of subsection \ref{nummethod} we have drawn the trajectories
of the effective energies given by Eqs. (\ref{eeff1}) and (\ref{eeff2}). 
The top frame of Fig. \ref{fig_eeff} shows short extracts of the complex trajectory $E_{\text{eff}}^{v=8,\text{ad}}(t)$ given in Eq. (\ref{eeff2}), deduced from the ideal adiabatic approximation of Eq. (\ref{eq:psiad}). In this case, only one Floquet eigenvector is present. Each ellipse represents the complex trajectory followed by $E_{\text{eff}}^{8,\text{ad}}(t)$ during one optical cycle, at three  intermediate times corresponding to given field parameters during the rise of the pulse, $E(t_1)=0.5E_{\text{max}}$, $E(t_2) = 0.75 E_{\text{max}}$ and $E(t_3) = E_{\text{max}}$. $E_{\text{max}}$ is the maximum field amplitude reached during the pulse.
The dotted line represents the trajectory followed by the first term of Eq. (\ref{eeff2}), i.e. the instantaneous Floquet eigenvalue $E_8\{ \epsilon(t) \}$. This is also the trajectory followed by the instantaneous centers of the ellipses resulting from the second term in Eq. (\ref{eeff2}). Four black circles also indicate the positions of $E_8\{ \epsilon(t) \}$ at four different times, $t_0=0$ (when there is no electric field) and at the intermediate times $t_1$, $t_2$, $t_3$ defined above. This dotted trajectory stays very close to the real axis because the control pulse is designed to follow a ZWR path. 
This frame should be compared with the middle and bottom frames of Fig. \ref{fig_eeff} which contains extracts of the effective energy trajectory $E_{\text{eff}}^{v=8}(t)$ obtained by using Eq. (\ref{eeff1}) with the exact solution of the TDSE. 
The ellipses have been recorded 
during short time intervals $[t_i,t_i+\Delta t]$ around the same intermediate times $t_1$, $t_2$, $t_3$,
with $\Delta t =$ 0.12 ps corresponding to a few tens of optical cycles (this explains the broadening of the curves). We have also shown the instantaneous center, i.e. the rolling mean over one optical cycle, $ \frac{1}{T(t)}  \int_t^{t+T(t)} E_{\text{eff}}^8(t') dt'$, during the whole pulse. 
If the adiabatic approximation is satisfied, this instantaneous center should be close of the eigenvalue trajectory (dotted line in the top frame). The more adiabatic is the dynamics, the closer from the top frame should be the trajectories. 

The middle frame has been obtained by using the pulse of Eq. (\ref{adiab_pulse1}) and (\ref{eq:enveloppe}), with the phase  $\theta(t) = \int_0^t \omega_{\text{eff}}^{\text{ZWR}}(t') dt'$. We observe a nice agreement with the ideal adiabatic trajectory shown in the top frame, with ellipsoids of growing radii whose centers closely follows  the instantaneous resonance trajectory. The deviations from the real axis are less than 0.2 cm$^{-1}$. 
As a comparison, the bottom frame shows the trajectories obtained by using the naive pulse of Eq. (\ref{eq:ins_pulse}) with the phase $\theta(t) =\omega_{\text{eff}}^{\text{ZWR}} \cdot t$. We observe very different results: Adiabaticity is rapidly lost, in agreement with the explanations developed in section \ref{sec3}. The deviation from adiabaticity is even more marked during the falling-off of the control field, leading to an erratic behavior in the instantaneous center trajectory. The chaotic path on the right side of panel (c) corresponds to the second half of the pulse and indicates that the adiabatic character is completely lost. These results are consistent with the direct population analysis of Fig.~\ref{adiab_ins}.

\begin{figure}
\begin{center}
\includegraphics[width=0.43\linewidth]{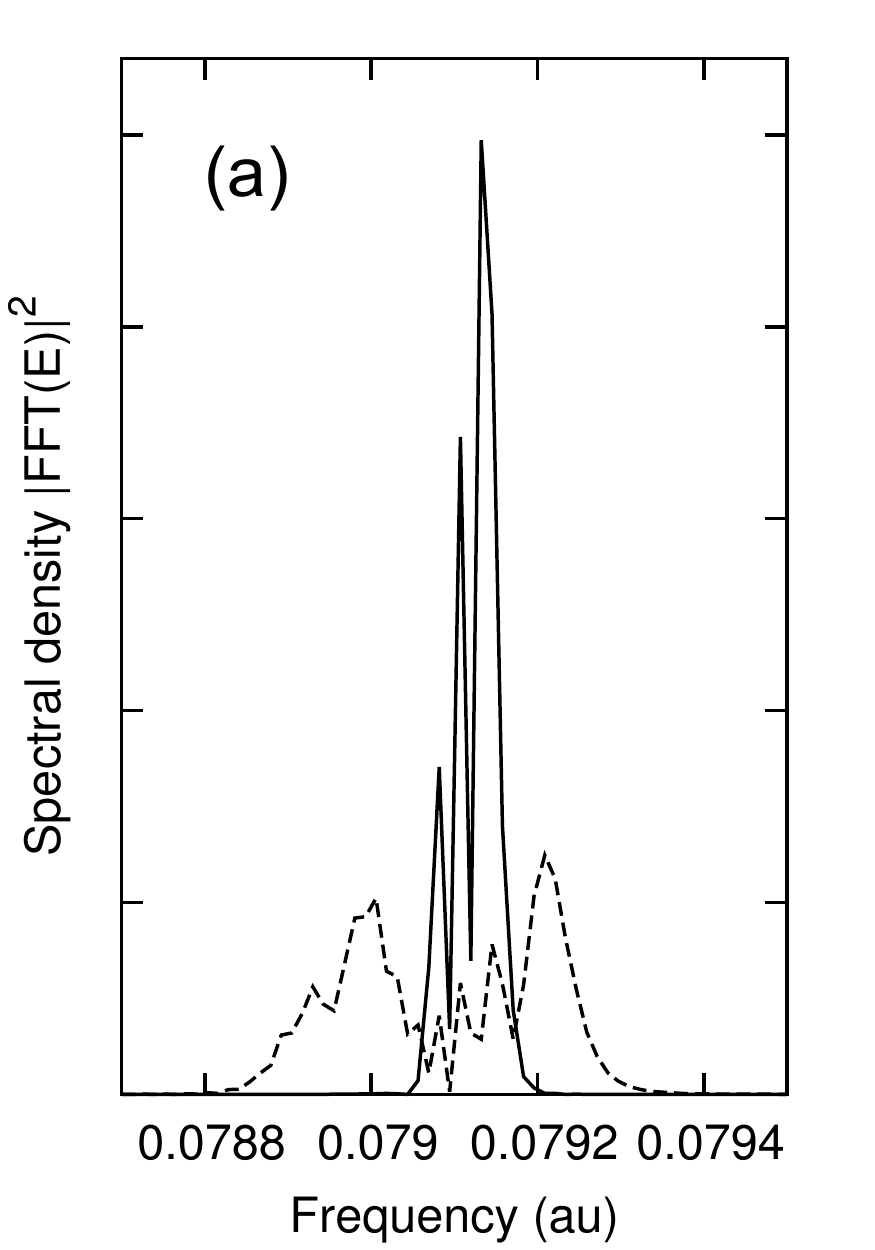}
\includegraphics[width=0.52\linewidth]{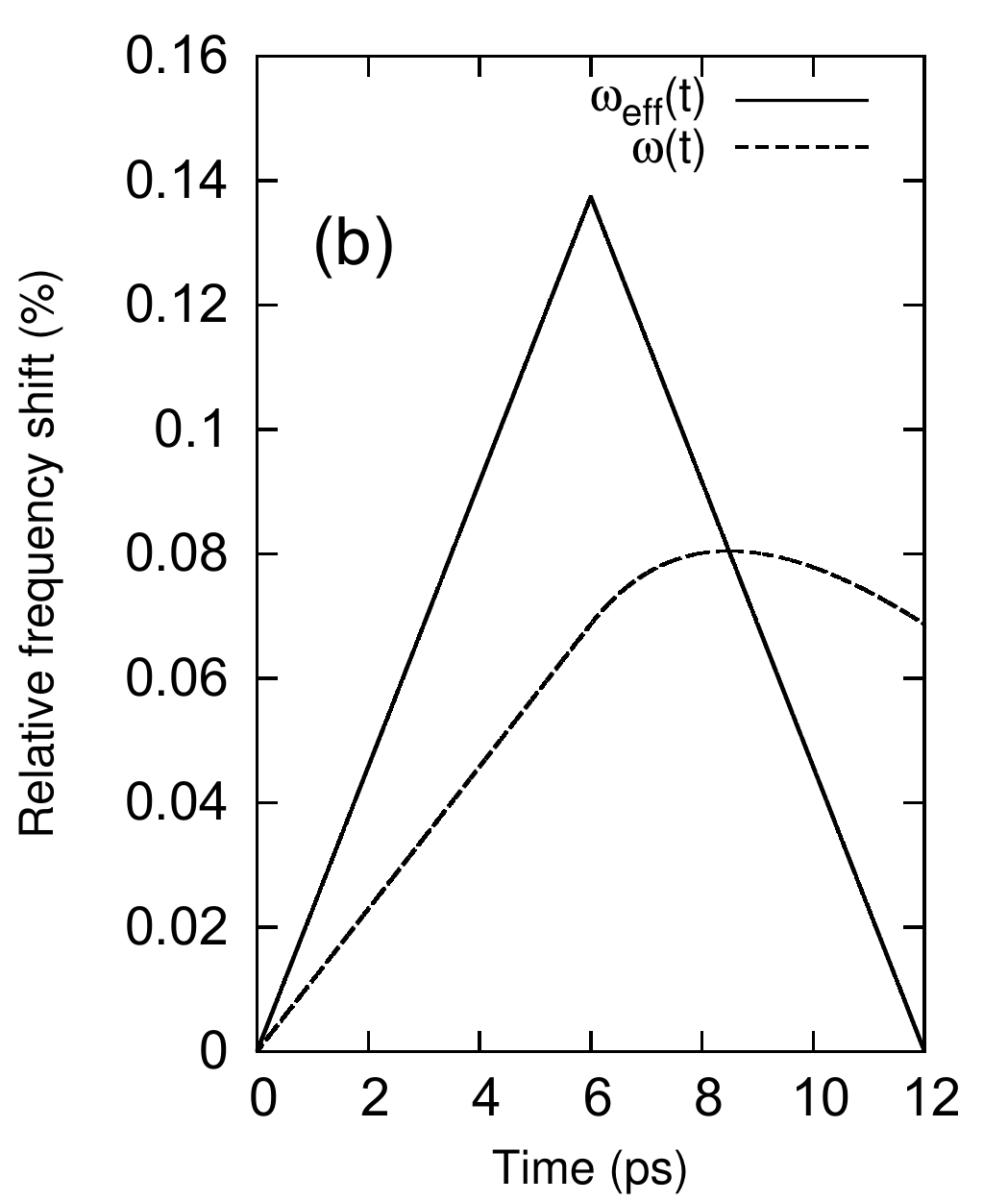}
\end{center}
\caption{Laser pulse properties: (a) Power spectrum of the laser pulse shaped using the adiabatic strategy of Eq. (\ref{adiab_pulse}), solid line, and the naive strategy of Eq. (\ref{eq:ins_pulse}), dashed line. 
(b) Frequency shift during the pulses, for the effective frequency
$ \frac{\omega_{\text{eff}}(t)-\omega(0)}{\omega(0)} $ 
(solid line) and the actual instantaneous frequency 
$ \frac{\omega(t)-\omega(0)}{\omega(0)} $ (dashed line). }
\label{fig:spectres}
\end{figure}

The results obtained using the two kind of pulses, corresponding to Eq. (\ref{eq:ins_pulse}) (naive pulse) or Eq. (\ref{adiab_pulse}) (adiabatic strategy), are very different, only the adiabatic strategy giving rise to an efficient protection against dissociation of the selected state. The two pulses are compared in Fig. \ref{fig:spectres}. A direct comparison is difficult due to the very large number of oscillations (more than 6000 optical cycles), therefore we have shown the power spectrum of each pulse on the left panel, together with the relative frequency shifts occuring during the pulses (right panel). 
The spectrum corresponding to the pulse of Eq. (\ref{adiab_pulse}) displays three major peaks close to the central frequency (corresponding to lambda = 576nm), whereas the one resulting from Eq.(\ref{eq:ins_pulse}) covers a much larger frequency range. At the selected scale, the difference between the two pulses is clearly noticeable. 
On the right panel of Fig. \ref{fig:spectres}, we have also drawn the relative frequency shifts 
$ \frac{\omega_{\text{eff}}(t)-\omega(0)}{\omega(0)} $ (for the effective frequency) 
and 
$ \frac{\omega(t)-\omega(0)}{\omega(0)} $ (using the actual instantaneous frequency). 
It is worthwile noting that the shifts remain very small all along the pulses. This is certainly an advantage for a future experimental implementation. 
The various numerical results previously shown in Fig.~\ref{adiab_ins} and Fig.~\ref{fig_eeff} can be better understood by noting that a strong difference piles up during the pulse between the effective and the instantaneous frequencies, explaining why the dynamics progressively gets away from an optimal adiabatic dynamics if we use one frequency instead of the other. Using the adiabatic strategy of Eq. (\ref{adiab_pulse}) for shaping the pulse is thus very important.

\subsection{Optimal ZWR paths.}
As graphically shown in Fig. (\ref{loci_zwr}), there are several ZWR paths originating from $v=$8 and labeled by the corresponding quasi-degenerate upper adiabatic potential levels $v_+$=0,1,2,3 of the semiclassical analysis. The question which is addressed now is the choice of the path which better fulfills the requirement of the filtration strategy. More precisely, all of these paths are well adapted to protect $v=$8 population from decaying. 
But, due to the anharmonicity and the density of the vibrational levels, 
a couple of laser parameters ($\lambda^{\text{ZWR}},I^{\text{ZWR}}$)  which are well identified as producing a ZWR of $v=$8, can also be rather well suited in producing a ZWR originating from a neighboring vibrational state $v=$7, 9 or 10. In such a situation, when following this ZWR path, not only $v=$8 population is protected against dissociation as it should be, but the dissociation of neighboring levels could also be severely slowed down. As a consequence the filtration strategy would loose its efficiency. Only very long duration pulses would achieve the complete decay of $v=$7,9 and 10, inducing the unwanted effect of rotational degrees of freedom coming into play producing rotational heating. The best choice among ZWR paths targets the highest contrast that could be expected between the widths of the resonances originating from $v=$8 (ideally zero) on one hand, and from $v=$7,9,10 (largest possible values, that is shortest lifetimes) on the other hand. Once again the semiclassical analysis reveals helpful for such a choice. 
 \begin{figure}
 	\includegraphics[width=0.9\linewidth]{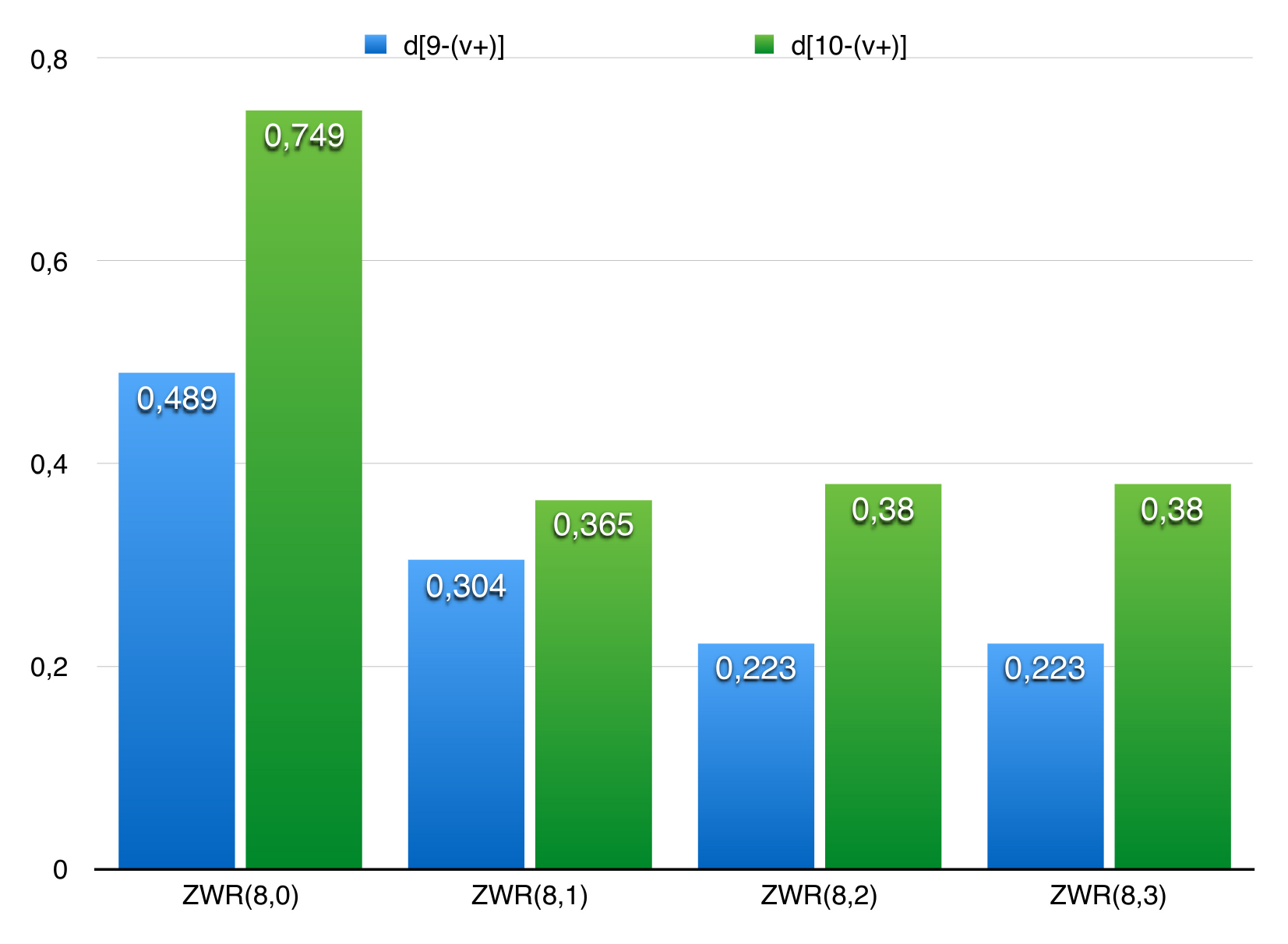}
 	\caption{Histograms displaying $d(9, v_++1, w)$ (in blue) and $d(10, v_++2, w)$ (in green) for the different ZWR paths ($v=8$, $v_+=0,1,2,3$) of Fig. (\ref{loci_zwr}). For the definition of $d$, see Eq.(\ref{eq:d_def}).}
 	\label{histogram}
 \end{figure}

 As indicated in Eq.(\ref{eq:Gammasem}), the semiclassical estimate of resonance widths is proportional to the square of the difference $(\varepsilon_{\tilde{v}}-\varepsilon_{\tilde{v}_{+}})^2$ of the energies of the corresponding levels accommodated by the adiabatic potentials $V_{\pm}(R)$. They are obtained by the numerical solution of Eqs.(\ref{eq:semiclassical1},\ref{eq:semiclassical2}) including the phase correction $\chi$. For field intensities close to zero, $\varepsilon_{\tilde{v}}$ is nothing but the energy $\varepsilon_{v}$, the energy of the ground state vibrational level $v$, whereas $\varepsilon_{\tilde{v}_{+}}$ is the semiclassical estimate $\varepsilon_{v_+}$ of the upper adiabatic potential vibrational levels with a $\chi=-\pi/4$ phase correction. Restricting the analysis to $v=$8,9 and 10, the field free values of $\varepsilon_{v}$  are calculated once for all, with the results: $\varepsilon_{8} =-82923.86 cm^{-1}$, $\varepsilon_{9} = -82915.99 cm^{-1}$, $\varepsilon_{10} =- 82909.80 cm^{-1}$, the zero-energy being chosen as the energy of two infinitely separated Na$^+$ ions, as in~\cite{magnier}.
The values of $\varepsilon_{v_+}$ depend on the specific wavelength $\lambda$ appropriately shifting the $V_+(R)$ potential in a field-dressed picture. As an example, for $\lambda = 576.396 nm$ corresponding to the almost zero intensity field value of the ZWR ($v=8$, $v_+=0$) of Fig. (\ref{loci_zwr}), one gets: $\varepsilon_{v_+=0} = -82924.21 cm^{-1}$, $\varepsilon_{v_+=1}= -82912.96 cm^{-1}$, $\varepsilon_{v_+=2}= -82906.45 cm^{-1}$. Remaining in the almost zero-field regime, the different ZWR paths displayed in Fig. (\ref{loci_zwr}) correspond to a quasi coincidence of $\varepsilon_{8}$ with one of the $\varepsilon_{v_+}$ ($v_+=0,1,2,3$) depending on the laser wavelength. Due to important anharmonicities and relatively dense vibrational distribution (for $v\ge$8), it may happen that, for some wavelengths, in addition to the quasi-degenerate couple of levels ($v=$8 with $v_+$=0,1,2,3) other couple of levels ($v'\ne$8 with $v'_+$) be in close proximity, leading unwanted secondary trapping processes, acting against the efficiency of the filtration strategy.
 \begin{figure}
 	\includegraphics[width=0.9\linewidth]{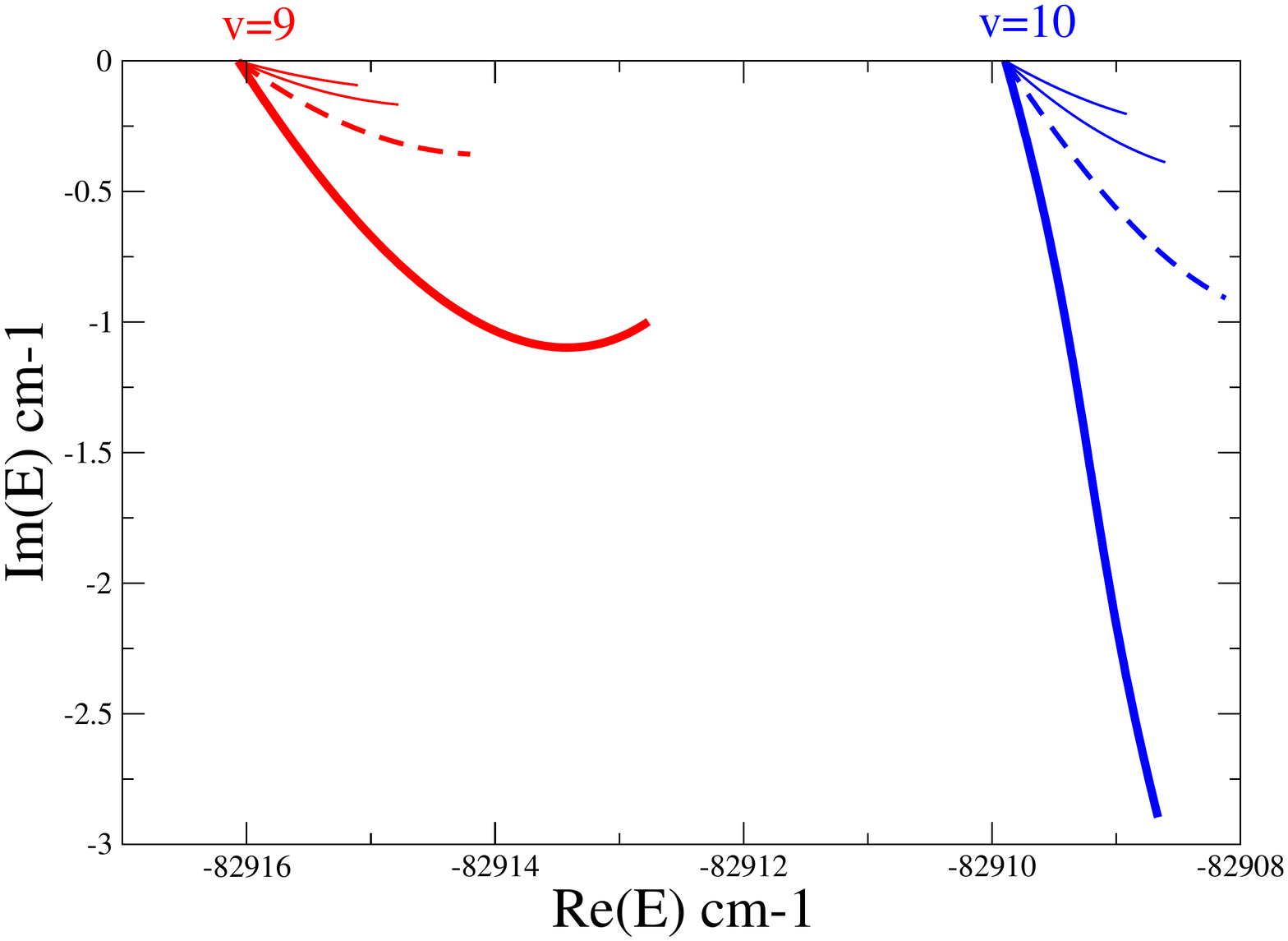}
 	\includegraphics[width=0.9\linewidth]{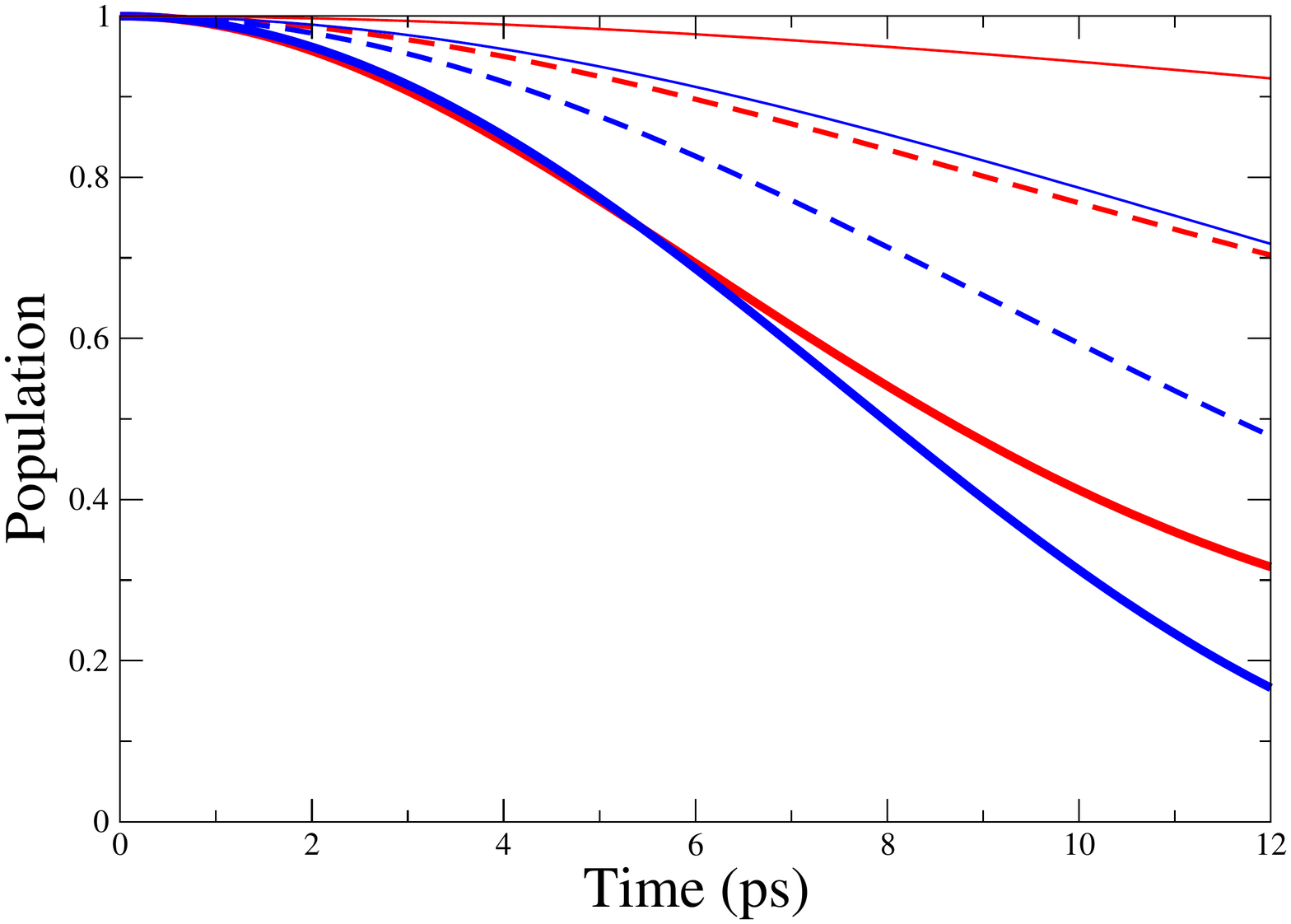}
 	\caption{(a) Resonance behaviors and decay rates in the complex energy plane with laser pulses shaped and chirped along the ZWR paths of Fig. (\ref{loci_zwr}).
 		(b) Estimate of the population decay as a function of time, using the adiabatic approximation of Eq. (\ref{P-undiss}).
 	Color code for both panels: $v=$9 in red and $v=$10 in blue, thick solid lines for ZWR(8, $v_+$=0), dashed lines for ZWR(8, $v_+$=1), thin solid lines for ZWR(8, $v_+$=2,3).}
 	\label{Gammas910}
 \end{figure}
 A quantitative analysis can be provided based on the definition, for a given wavelength, of a dimensionless parameter $d(v, v_+, w)$ which measures the energy proximity among a given couple $\varepsilon_{v}$ and $\varepsilon_{v_+}$:
 \begin{equation}
 d(v,v_+,w)=\frac{\varepsilon_{v}-\varepsilon_{v_+}}{\Delta \varepsilon_{w} }
 \label{eq:d_def}
 \end{equation}
 $w$ labels one of the two anharmonicity windows of the vibrational levels sequence, with an energy extension $\Delta \varepsilon_{w}=\varepsilon_{v+1}-\varepsilon_{v}$. More precisely, for a given $v$, $w=1$ depicts a situation where a given $\varepsilon_{v_+}$ is within the window $\Delta \varepsilon_{1}=\varepsilon_{v}-\varepsilon_{v-1}$ (for example $\varepsilon_{9}-\varepsilon_{8}=7.87 cm^{-1}$), whereas $w=2$ corresponds to $\Delta \varepsilon_{2}=\varepsilon_{v+1}-\varepsilon_{v}$ (for example $\varepsilon_{10}-\varepsilon_{9}=6.19 cm^{-1}$). Obviously $d$ may vary between 0 (exact degeneracy $\varepsilon_{v}=\varepsilon_{v_+}$) to 1. In practical calculations done for the different ZWR paths originating from $v=$8, we obtain semiclassical estimates for $d(8, v_+=0,1,2,3, w=1,2)$ typically less than 0.04. The additional important information emerges from the analysis of $d(9,v_++1, w)$ and $d(10,v_++2, w)$ addressing the following question: When a given $v_+$ is almost degenerate with $v=$8, how much is ($v_++1$) close to $v=9$, or ($v_++2$) to $v=10$? The histograms of Fig.(\ref{histogram}), displaying the corresponding values of $d$, for each of the ZWR paths, answer this question and help in the choice of the best adapted path leading to the maximum contrast, that is for an almost zero estimate for $d(8, v_+, 1)$, the largest possible values for $d(9, v_++1, 2)$ and $d(10, v_++2, 3)$, in order for the filtration be the most efficient. This is obviously obtained for ZWR($v=8$, $v_+=0$) on which the filtration will ultimately be based.
This choice is also confirmed by looking at the Floquet results displayed in Fig. (\ref{morphology_zwr}.b). There are three local minima for the resonance width $\Gamma_8$. The third minimum (wavelenght around 576.4$nm$), which corresponds to the semi-classical coincidence of $(v=8,v_+=0)$, is the one which faces simultaneously the two best-marked maxima for both $\Gamma_9$ and $\Gamma_{10}$. The contrast is then optimal in comparison to the two other possible minima of $\Gamma_8$ appearing around $\lambda=569 nm$ and $\lambda=572.5 nm$.

 This semiclassical analysis conducted on near zero-field limit, can be confirmed by the time-dependent quantum behavior of resonance widths during their adiabatic transport all along the ZWR paths of Fig.(\ref{loci_zwr}). The result is displayed in Fig.(\ref{Gammas910}a) for resonances originating from $v=9$ and $v=10$ with laser pulses shaped and chirped as to follow the ZWR paths ($v=8$, $v_+=0,1,2,3$). A much more pronounced decay rate (larger imaginary parts of the energies) is observed when following ZWR ($v=8$, $v_+$=0).
 Within a purely adiabatic hypothesis of a single (not contaminated) resonance transport, these rates can be used to calculate the resulting vibrational population decays during the pulse through Eq.(\ref{P-undiss}). The results are gathered in Fig.(\ref{Gammas910}b) which clearly shows that the most important $v=$9 and 10 populations depletion are observed when the transport is along the ZWR ($v=8$, $v_+$=0) path. This is a complete confirmation of the semiclassical choice of this particular path as producing the highest possible decay of $v=$9, 10 populations, while quenching the one of $v=$8.

 \begin{figure}
 	\includegraphics[width=0.9\linewidth]{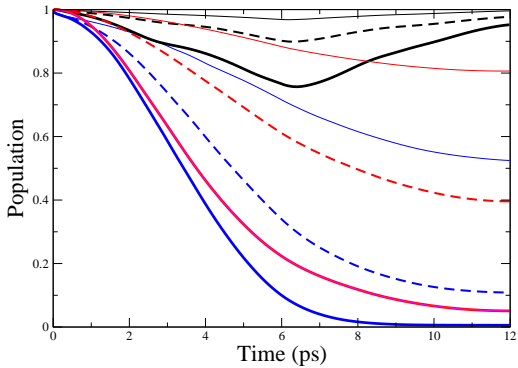}
 	\caption{Vibrational populations as a function of time for three different laser pulses, adiabatically following the ZWR paths of  Fig.(\ref{loci_zwr}). 
 	Thick solid lines for the path corresponding to ZWR ($v=8, v_+=0$), dashed lines for the path ($v=8, v_+=1$), thin solid lines for the path ($v=8, v_+=2$). Color code: Black for $v=$8, red for $v=$9, blue for $v=$10. }
 	\label{populations}
 \end{figure}

 The implementation of the filtration scheme, within the hypothesis of an initial ensemble of vibrational states, or of an initial coherent vibrational superposition, is summarized in Fig.(\ref{populations}). We assume an equally populated field-free initial vibrational distribution among levels $v=8$ (in black), $v=9$ (in red) and $v=10$ (in blue). Three laser pulses are considered, which adiabatically follow the ZWR originating from $v=8$ as depicted in Fig.(\ref{loci_zwr}) and shaped such as to quench $v=8$ dissociation. ZWR ($v=8, v_+=0$) is indicated by thick solid lines, ZWR ($v=8, v_+=1$) by dashed lines, and ZWR ($v=8, v_+=2$) by thin solid lines. 
Full quantum dynamical simulations based on wavepacket propagation are used for solving the TDSE Eq.(\ref{TDSE}) as explained in section \ref{nummethod}. 
The first observation is that for all ZWR paths the initial population of $v=$8 is well protected against dissociation, as is expected from an adiabatic transport. The final $v=$8 undissociated populations are all above 95$\%$, which remains a remarkable result for the ZWR quenching mechanism itself. The second important information concerns the overall filtration process efficiency. While for ZWR ($v=8, v_+=2$) a large remaining population on $v=$9, 10 levels (more than 80$\%$ and 50$\%$ respectively) renders the filtration inefficient, only the choice of ZWR ($v=8$, $v_+=0$) leads to acceptable results, with less than 10$\%$ of undissociated population on $v=9, 10$. The vibrational cooling control is achieved as the system is left on a single vibrational level, namely $v=8$. A final STIRAP process \cite{vitanov2001} could optionally be referred to for transferring the population from $v=8$ to $v=0$.

\section{Conclusion}\label{sec5}
We propose a laser control strategy for the vibrational cooling of diatomic molecules based on an efficient filtration scheme.The basic mechanism which is referred to is a destructive interference between outgoing fluxes from the two field-dressed electronic states involved in the photodissociation process. In principle, infinitely long lived resonances (ZWRs) result from such interference, for some well tuned couples of laser parameters (wavelength and intensity). Starting from an initial vibrational population distribution, filtration aims at a selective decay among them; namely, all levels, except one, should dissociate. The proposed scheme goes through the optimization of a laser pulse envelop shaped and frequency chirped in such a way as to protect against dissociation a given initial vibrational level $v$. This is achieved by an adiabatic transport of the $v$ population on its corresponding ZWR which is continuously followed all along the pulse duration. Two important issues that turn out to be crucial for the success of the process when dealing with heavy systems like Na$_2$, and that have not been addressed previously, are emphasized in this work.

The more fundamental first one concerns the challenging problem of adiabaticity in decaying system dynamics involving a dissociative continuum. Actually this is solved by adapting the adiabatic Floquet approach (originally derived for bound states) to the special case of ZWRs, taking advantage of their non-decaying peculiarity, although being coupled to a dissociative continuum (BICs, Bound States in Continuum). It is only through such an adiabatic transport that we can protect the population of a vibrational level in a robust way. 

The second issue concerns more specifically the application in mind. We have shown that the family of ZWRs, originating from the field-free initial vibrational level $v$ to be protected, is organized in terms of several paths in the laser parameter plane. If all the members of this family  are convenient to quench the dissociation from $v$,  some of the laser pulses built to follow such paths may have characteristics close to be appropriate for partly protecting neighboring vibrational populations as well. This situation arises when transposing the control scheme to heavier diatomics like Na$_2$, basically due to anharmonicity and relatively high vibrational levels density.  Its occurrence is obviously against the efficiency of the filtration process. A semiclassical analysis developed in detail helps the choice of the pulse offering the best compromise between quenching $v$ and providing highest possible decay rates for the neighboring vibrational states $v'$ ($v'\ne v$) .

A full quantum wavepacket propagation using the optimal pulse confirms the expectations of both the adiabatic Floquet theory and the semiclassical analysis. In the hypothesis of an initial equal vibrational population partition on $v=$8,9, and 10 for Na$_2$ prepared by photoassociation, $v=$8 is protected up to 95$\%$ of its initial population, whereas only less than 5$\%$ of the $v=$9 and 10 populations are left. It is worthwhile noting that such a control is achieved with a total pulse duration not exceeding 12$ps$, thus avoiding rotational heating of the molecule. Our final claim is that the proposed laser pulse characteristics, namely rather modest frequency chirp amplitudes around $\lambda= 570nm$, moderate intensities (less than $10^{10}W/cm^2$), realistic pulse duration (less than 12$ps$), remain experimentally feasible. Moreover, within the adiabatic transport frame that has been developed, the model becomes finally generic enough to be transposed  to the vibrational cooling of other alkali dimers.

\begin{acknowledgments}
O. A. acknowledges support from the European Union (Project No. ITN-2010-264951, CORINF). 
Part of the simulations have been executed on computers of the Utinam Institute of the Universit\'e de Franche-Comt\'e, supported by the R\'egion de Franche-Comt\'e and Institut des Sciences de l'Univers (INSU).
\end{acknowledgments}

\bibliographystyle{apsrev}

\end{document}